\documentclass[journal]{IEEEtran}
\ifCLASSINFOpdf
\else
\fi
\usepackage{graphicx,multirow,lscape,color,times,url,amsfonts,amsmath,array}

\usepackage{amsfonts}
\usepackage{amsmath}
\usepackage{amssymb}
\usepackage{multirow}
\usepackage{epsfig}
\usepackage{graphpap}
\usepackage{eurosym}
\usepackage{algorithm}
\usepackage{algpseudocode}
\usepackage{color}
\usepackage[noadjust]{cite}
\newtheorem{lem}{Lemma}
\newtheorem{rem}{Remark}
\usepackage{graphpap}
\usepackage{balance}
\usepackage{mathtools}
\usepackage[font=footnotesize,labelfont=scriptsize,justification=centering]{caption}
\usepackage{bm}
\usepackage[
singlelinecheck=false
]{caption}
\usepackage[T1]{fontenc}
\usepackage{color, colortbl}
\definecolor{Gray}{gray}{0.9}
\definecolor{LightGray}{gray}{0.7}
\definecolor{LightCyan}{rgb}{0.88,1,1}
\usepackage[first=0,last=9]{lcg}
\newtheorem{proposition}{Proposition}
\begin{document}
\title{Coexistence of MIMO Radar and FD MIMO Cellular Systems with QoS Considerations}
\author{\IEEEauthorblockN{Sudip Biswas, Keshav Singh, Omid Taghizadeh, and Tharmalingam Ratnarajah}
\thanks{\hrulefill}
\thanks{Manuscript received January 03, 2018; revised March 24, 2018; accepted August 03, 2018. The associate editor coordinating the review of this paper and approving it for publication was Prof. Shi Jin.}
\thanks{This work was supported by the U.K. Engineering and Physical Sciences Research Council (EPSRC) under Grant EP/N014073/1 and the UK-India Education and Research Initiative Thematic Partnerships under grant number DST-UKIERI-2016-17-0060.}
\thanks{S. Biswas, K. Singh, and T. Ratnarajah are with the Institute for Digital Communications, School of Engineering, University of Edinburgh, Edinburgh, UK (e-mail:\{k.singh, sudip.biswas, t.ratnarajahg\}@ed.ac.uk).}
\thanks{O. Taghizadeh is with the Institute for Theoretical Information Technology, RWTH Aachen University, Germany (e-mail: taghizadeh@ti.rwth-aachen.de).}
\thanks{The corresponding author of this paper is Keshav Singh.}
}
\maketitle
\begin{abstract}
In this work, the feasibility of spectrum sharing between a multiple-input multiple-output (MIMO) radar system (RS) and a  MIMO cellular system (CS), comprising of a full duplex (FD) base station (BS) serving multiple downlink and uplink users at the same time and frequency is investigated. While a joint transceiver design technique at the CS's BS and users is proposed to maximise the probability of detection (PoD) of the MIMO RS, subject to constraints of quality of service (QoS) of users and transmit power at the CS, null-space based waveform projection is used to mitigate the interference from RS towards CS. In particular, the proposed technique optimises the performance of PoD of RS by maximising its lower bound, which is obtained by exploiting the monotonically increasing relationship of PoD and its non-centrality parameter. Numerical results show the utility of the proposed spectrum sharing framework, but with certain trade-offs in performance corresponding to RS's transmit power, RS's PoD, CS's residual self interference power at the FD BS and QoS of users.
\end{abstract}


\section{{Introduction}}
{The proliferation of wireless devices and services along with static spectrum allocation have resulted in the dearth of spectral resources, which has empowered the vision of spectrum sharing between federal incumbents, such as radar (maritime, surveillance, weather, etc.) and commercial communication systems, such as cellular systems (CSs) \cite{FCC02,FCC12}. As a result, coexistence between radar and communication systems through technologies, such as Cognitive Radios (CRs)~\cite{Haykin05}, Licensed/Authorized Shared Access (LSA/ASA) \cite{CEPT14, Matinmikko14} and Spectrum Access Systems (SAS)~\cite{NTIA10} has captured the attention of both academia and industry. However, irrespective of the technology being used, spectrum sharing between radar and communication systems brings a new set of challenges into picture, such as the \textit{harmful interferences generated by} \textit{CSs towards the radar} and vice-versa, which can potentially degrade the quality-of-service (QoS) of both systems. }

In light of this, while the authors in \cite{Marcus09} addressed the possibility of spectrum sharing between a rotating radar and CS, the problem of spectrum sharing between a pulsed, search radar (primary) and $802.11$ WLAN (secondary) was investigated in \cite{Hessar15}. Similarly, an opportunistic spectrum access approach was proposed in \cite{Saruthirathanaworakun12} for spectrum sharing between rotating radar and CS. In this approach, the communication system is permitted to transmit signals \textit{if and only if}  the space and frequency spectra are not being utilised by the radar, hence prohibiting simultaneous operation of the radar and communication system. To alleviate this issue, LSA/ASA was studied in \cite{CEPT14, Matinmikko14}, which allows for the channel state information (CSI) of both systems to be accessible to each other to an extent, where by techniques, such as null-space projection (NSP) of waveforms \cite{Babaei13}, beamforming \cite{Liu18}, etc., can be utilised to mitigate interference at the coexisted systems. Furthermore, to increase waveform diversity and attain higher detection probability, the rotating radar was extended to multiple-input multiple-output (MIMO) radar in \cite{Khawar15}--\cite{Deng13}. Accordingly, while a MIMO radar operating in the presence of clutter and a MIMO CS was considered in \cite{Li16}, a null-space projection (NSP) technique for waveform design was proposed in \cite{Khawar15} to facilitate the coexistence between a MIMO radar and a MIMO CS. Further, a signal processing scheme for a MIMO radar was proposed in \cite{Deng13} for effectively minimising the arbitrary interferences generated by CSs from any direction towards the radar.

Besides spectrum sharing, full-duplex (FD) is another emerging technology that can potentially double the spectrum efficiency of currently deployed half duplex (HD) systems by fully utilising the spectrum resources in the same time and frequency \cite{Biswas16,Biswas_tcom2016}. However, the self-interference (SI) generated by  signal leakage from the transmitting antennas to its receiving antennas, dominates the performance of FD systems. However, recent advancements in interference cancellation techniques~\cite{duarte2010} suggest that SI can be mitigated in the analog and digital domain to the extent that only a residual SI (RSI) is left behind. This RSI is caused by  the non-ideal nature of the transmit and receive chains~\cite{li2014}, also known as hardware impairments (amplifier non-linearity, phase noise, I/Q channel imbalance, etc.) and can be mitigated through digital beamforming techniques. Another predicament of FD systems is the the co-channel interference (CCI) to the downlink (DL) users caused by the transmission from the uplink (UL) users, which has also been shown in literature to be suppressed through digital beamforming and user-clustering techniques. Accordingly, to harvest the full potential of FD, in recent studies~\cite{Biswas_tcom2016,nguyen2014,Cirik_tvt_2017} RSI and CCI were integrated in the design of transceivers for FD systems. However, most prior works consider perfect channel state information (CSI) at the transmitter, which is improbable to attain due to large training overhead and low signal-to-noise-ratio obtained after beamforming. Hence, it is  important to design robust transceivers, which can provide respectable performance for the co-existed systems even with imperfect or limited CSI estimates.

{Motivated by the above discussion, in this paper {we consider a two-tier spectrum sharing framework, where a multiple-input-multiple-output (MIMO) radar system (RS) is the spectrum sharing entity and a FD MIMO CS is the beneficiary.} Unlike existing literature \cite{Marcus09}--\cite{Deng13}, in this work we consider a framework wherein, by utilising the spectrum of the MIMO RS, the FD CS's BS serves multiple downlink (DL) and uplink (UL) users simultaneously in same time and frequency resources. At this point, we would like to note that though the consideration of FD mode at the CS provides better QoS for the users, it also induces {\textit{higher intensity of interference}} {\textit{towards the RS, which might affect its probability of detection (PoD)}}. Hence, the requirements for the coexistence of a FD CS, with a MIMO RS requires revisiting. Further, the inclusions of  imperfect CSI, hardware impairments and CCI in the design of the proposed co-existed network result in a rather arduous problem, which requires rigorous optimisation and analysis and is addressed in this paper.
{Accordingly, to enable efficient spectrum sharing between a FD MU MIMO CS and MIMO RS, we focus on the design of:} 
\begin{itemize}
			\item {Beamforming matrices at the FD CS to mitigate interference towards the RS, while also providing data to its users. In particular, we formulate an optimisation problem for maximising the PoD of the RS for a given worst-case channel set, subject to the constraints of QoS at the UL and DL users and powers at the UL users and the FD BS.}
			\item  {Projection matrix at the MIMO RS to mitigate interference towards the FD CS. In particular, all the interference channels from the RS to CS are combined into one and a null-space is created. The RS's signal is then projected onto the null-space of the combined interference channel.}			
\end{itemize}}

The following notations are used throughout this paper. Boldface capital and small letters denote matrices and vectors, respectively. While the transpose and the conjugate transpose are  denoted by $(\cdot)^{T}$ and $(\cdot)^{H}$, respectively, $\|\textbf{A}\|_{F}$ and $\|\textbf{a}\|_{2}$ denote the Frobenius norm of a matrix $\textbf{A}$ and the Euclidean norm of a vector $\textbf{a}$, respectively. $\otimes$ denotes  Kronecker product and $\perp$ denotes the statistical independence. The matrices ${\bf I}_N $ and ${\bf 0}_{M \times N}$ denote a $N \times N$ identity matrix and a $M \times N$ zero matrix, respectively. The notations $\mathbb E{(\cdot)}$ and $\text{tr}(\cdot)$ refer to expectation and trace, respectively and $diag(\textbf{A})$ and  $vec(\textbf{A})$ generate a diagonal matrix with the same diagonal element as $\textbf{A}$ and a $mn\times1$ column vector obtained by stacking the columns of the matrix $\textbf{A}$ on top of one another, respectively.

\section{{System Model}}\label{section_system_model}
We consider the coexistence of a FD multi-user MIMO CS with a MIMO radar system (RS) as shown in Fig. \ref{system_model}, where the CS operates in the spectrum shared by the  RS over a bandwidth of $B$ Hz. The CS comprises of a FD MIMO BS, which consists of $M_0$ transmit and $N_0$ receive antennas, and $J$ DL, and $K$ UL users. All DL and UL users operate in half-duplex (HD) mode and each DL and UL user is equipped with $N_j$ receive and $M_k$ transmit antennas, respectively. Furthermore, the MIMO RS is equipped with $R_R$ receive and $R_T$ transmit antennas. In the following sub-sections, we provide details on the architecture of the two systems. 
\subsubsection{FD MIMO CS} 
\begin{figure}[t!]
\centering
\includegraphics[scale=0.33, angle=270]{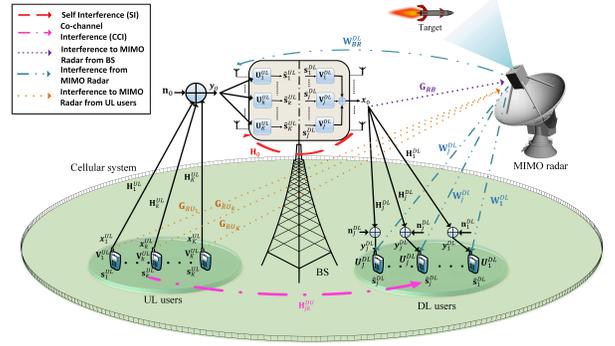}
\caption{Spectrum coexistence between FD MIMO CS and MIMO RS.}
\label{system_model}
\end{figure}

As shown in Fig. \ref{system_model}, $\mathbf{H}_{k}^{UL} \in \mathbb{C}^{N_0 \times M_k}$ and $\mathbf{H}_{j}^{DL} \in \mathbb{C}^{N_j \times M_0}$ represent the $k$-th UL and the $j$-th DL channel, respectively. The SI channel at the FD BS and the CCI channel between the $k$-th UL and $j$-th DL users are denoted as $\mathbf{H}_{0} \in \mathbb{C}^{N_0 \times M_0}$ and $\mathbf{H}_{jk}^{DU} \in \mathbb{C}^{N_j \times M_k}$, respectively. Finally, the \textit{interference channels} from the FD BS and $k$th UL user to the RS are denoted as $\mathbf {G}_{RB}\in\mathbb{C}^{R_R\times M_0}$ and $\mathbf {G}_{RU_k}\in\mathbb{C}^{R_R\times M_k}$, respectively.
Now, let $\mathbf{s}_k^{UL}(l) \in \mathbb{C}^{ d_k^{UL} \times 1}$ and $\mathbf{s}_j^{DL}(l)\in \mathbb{C}^{ d_j^{DL} \times 1}$ denote the transmitted symbols by the the $k$-th UL user and the FD BS, respectively, at time index $l$, with $l=1,\dots , L$, such that $\mathbb{E}\left[\mathbf{s}_k^{UL} (l)\left( \mathbf{s}_k^{UL}(l)\right)^H \right] =\mathbf{I}_{d_k^{UL}}$ and $\mathbb{E}\left[\mathbf{s}_j^{DL}(l)\left( \mathbf{s}_j^{DL}(l)\right)^H \right] =\mathbf{I}_{d_j^{DL}}$. { Here, $d_k^{UL}$ and $d_j^{DL}$ denote the data streams from the $k$th UL user and for the $j$th DL user, respectively, and $L$ is the total number of time samples used for CS's communication. } The symbols $\mathbf{s}_k^{UL}$ and $\mathbf{s}_j^{DL}$ are first precoded by matrices $\mathbf{V}_k^{UL} \in \mathbb{C}^{M_k \times d_k^{UL}}$ and $\mathbf{V}_j^{DL} \in \mathbb{C}^{M_0 \times d_j^{DL}}$, respectively, such that the signals transmitted from the $k$-th UL user and the FD BS at time index $l$ are given as $\mathbf{x}_k^{UL}(l)= \mathbf{V}_k^{UL}\mathbf{s}_k^{UL}(l)$ and $\mathbf{x}_0(l)= \sum_{j=1}^J \mathbf{V}_j^{DL}\mathbf{s}_j^{DL}(l)$, respectively. Further, similar to \cite{day2012}, we model the imperfections in the transmitter/receiver chains (oscillators, analog-to-digital converters (ADCs), digital-to-analog converters (DACs), and power amplifiers) as an additive Gaussian noise. 
In particular, we consider an additive white Gaussian term as ``transmitter noise'' (``receiver distortion'') at each transmit (receive) antenna, whose variance is $\psi(\upsilon)$ times the power of the undistorted signal at the corresponding chain.

\subsubsection{MIMO RS}
Let $\mathbf{W}_i$ denote the \textit{interference channels} shared by the MIMO RS with the CS, where $i=1,\dots,\mathcal{I}$, with $\mathcal I=1+J$ . As shown in Fig. \ref{system_model}, $\mathbf{W}_{BR}^{DL}\in \mathbb{C}^{N_0 \times R_T}$ and $\mathbf{W}_j^{DL} \in \mathbb{C}^{N_j \times R_T}$ denotes  the interference channels from RS's transmitter to CS's BS and $j$-th DL user, respectively, with $\{\mathbf{W}_{BR}^{DL},\mathbf{W}_{j}^{DL}\}\subseteq\mathbf{W},\,\forall j$. Let $\mathbf{s}_R(l) \in \mathbb{C}^{R_T\times 1}$ represents the transmitted symbol by the RS  at time index $l$, with
 $l=1,\dots,L_R$, and $\mathbb{E}\left[\mathbf{s}_R(l)\left(\mathbf{s}_R(l)\right)^{H}\right]=\mathbf{I}_{R_T}$. {Here, $L_R$ is the total number of time samples used for the RS's communication.}
 For the ease of derivation, hereinafter we assume that the time duration of the RS's waveform is the same as the communication signals with $L_R=L$ \cite{Liu18}.
\subsubsection{CSI acquisition} Acquiring CSI at both systems is important to ensure an interference free communication. {In this work, we assume  that some  amount of CSI, if not full is available at the communication nodes\footnote{CSI estimation can be performed via the exchange of  training sequences and feedback, and the application of usual CSI estimation methods \cite{Biswas_tcom2016}.}.} For the CS, providing its CSI to the RS is incentivised  by the promise of zero interference from the RS. 
On the other hand, it is more challenging to obtain an accurate estimate of the CSI of the RS at the CS, as the RS might not be willing to cooperate with the CS owing to security concerns. Hence, it might not be possible to obtain a full CSI at the CS and only partial CSI may be obtained through techniques such as blind environmental learning~\cite{gharavol2010}, realisation of a band manager with the authority of exchanging CSI between the  RS and CS~\cite{ban2009}, etc. Hence, to model the imperfections caused due to imperfect CSI, in this paper we will consider a norm-bounded channel estimation error model for the links between the CS and RS. 

\section{{Spectrum Sharing MIMO RS}}
As discussed in Section I, the considered sharing framework requires both the RS and CS to work in tandem to mitigate interference towards each other. Accordingly, in this section, we formulate the part played by the MIMO RS for efficient spectrum sharing between the two systems. The goal of the RS is to map $\mathbf{s}_R(l)$ onto the null-space of $\mathbf{W}$ in order to avoid interference towards the CS. 
Accordingly, the symbols $\mathbf{s}_R$ are first multiplied by a projector matrix $\mathbf{P}\in\mathbb{C}^{R_T\times R_T}$, such that the signal transmitted by the RS is given by $\mathbf{x}_R(l)={\mathbf{P}}\mathbf{s}_{R}(l)$.
\subsubsection{Target detection}In order to detect a target in the far field, we apply a binary hypothesis and choose between two cases: 1) $\mathcal{H}_0$: No target but active CS and 2) $\mathcal{H}_1$: Both the target and the CS are active. Further, in this work we consider only a single target with no interference from other sources, but the CS in order to study the impact of CS's interference on PoD of RS. 

Accordingly, the hypothesis testing problem for an echo wave in a single range-Doppler bin of the RS can be written as\footnote{For simplicity, we ignore the interference from clutter and false targets.} in~\eqref{hypothesis} as shown on the top of the next page,
\begin{figure*}
 \begin{equation}\label{hypothesis}
\textbf{y}_{R}(l) =
    \begin{cases}
      \mathcal{H}_1:\alpha\sqrt{P}_{R}\textbf{A}\left(\phi\right)\mathbf{P}\textbf{s}_{R}(l)+\overbrace{\textbf{G}_{RB}\left(\sum\nolimits_{j=1}^J \mathbf{V}_j^{DL}\mathbf{s}_j^{DL}(l)+\mathbf{c}_0\right)}^{\text{Interference from BS}}\\
      \hspace{15em}+\underbrace{\sum\nolimits_{k=1}^{K}{\mathbf{G}_{RU_{k}}}\left(\mathbf{V}_k^{UL}\mathbf{s}_k^{UL}(l)+\mathbf{c}_{k}^{UL}\right)}_{\text{Interference from UL users}}+\textbf{n}_{R}(l), & 1\leq l\leq L,\\
      \mathcal{H}_0:\textbf{G}_{RB}\left(\sum\nolimits_{j=1}^J \mathbf{V}_j^{DL}\mathbf{s}_j^{DL}(l)+\mathbf{c}_0\right)+\sum\nolimits_{k=1}^{K}{\mathbf{G}_{RU_{k}}}\left(\mathbf{V}_k^{UL}\mathbf{s}_k^{UL}(l)+\mathbf{c}_{k}^{UL}\right)+\textbf{n}_{R}(l), & 1\leq l\leq L,
    \end{cases}      
\end{equation} 
\hrule
\end{figure*}
where $\mathcal{H}_1$ comprises of the discrete time signal vector received by RS at an angle $\phi$, and $\mathcal{H}_1$ comprises of the interference plus noise signals from the CS only. The symbols $\mathbf{c}_0$ and $\mathbf{c}_{k}^{UL}$ are defined in Section IV. In the above, $P_{R}$ denotes the transmit power of RS, $\alpha_r$ indicates the complex path loss exponent of the radar-target-radar path including the propagation loss and the coefficient of reflection, $\textbf{n}_{R}(l)$ $\sim\mathcal{CN}\left(0,\sigma_{R}^{2}\textbf{I}_{R}\right)$, and $\textbf{A}\left(\phi\right)$ denotes the transmit-receive steering matrix defined as
 \begin{align}\label{Tx_Rx_steering_matrix}
\textbf{A}\left(\phi\right)\triangleq \textbf{a}_{R}\left(\phi\right)\textbf{a}_{T}^{T}\left(\phi\right).
	\end{align}
Here, $\textbf{a}_{T} \in \mathbb{C}^{R_{T} \times 1}$ and $\textbf{a}_{R} \in\mathbb{C}^{R_{R} \times 1}$ denote the transmit and receive steering vectors of RS's antenna array. Hereinafter, we assume that $R_{R}=R_{T}=R$ and $\textbf{a}_{R}\left(\phi\right)=\textbf{a}_{T}\left(\phi\right)=\textbf{a}\left(\phi\right)$. Accordingly, the $i$-th element at the $r$-th column of the matrix $\textbf{A}(\phi)$, can be written as
 \begin{align}
{A}_{ir}\left(\phi\right)=\text{exp}\left(-j\dfrac{2\pi}{\lambda}\left[\text{sin}\left(\phi\right);\text{cos}\left(\phi\right)\right]^{T}\left(\textbf{z}_{i}+\textbf{z}_{r}\right)\right),
	\end{align}	
where $\textbf{z}_{i}=\left[z_{i}^{1};z_{i}^{2}\right]$ is the location of the $i$-th element of the antenna array and $\lambda$ is the wavelength of the carrier. 
 Since, the deterministic parameters $\alpha_r$ and $\phi$ are unknown\footnote{Due to the availability of CS's CSI at the RS, we assume that the covariance matrix of the interference-plus-noise has been accurately estimated by the RS.}, we adopt the generalised likelihood ratio test (GLRT)~\cite{Xu07}, which has the advantage of replacing the unknown parameters with their maximum likelihood (ML) estimates for determining the probability of detection (PoD). Hereinafter, unless otherwise stated, we drop the time index $l$ in this section for notational convenience. The sufficient statistic of the received signal can then be found using matched filtering as \cite{Bekkerman06}
 \begin{align}\label{Y_hat}
\hat{\textbf{Y}}=&\dfrac{1}{\sqrt{L}}\sum\nolimits_{l=1}^{L}\textbf{y}_{R}\mathbf{x}_R^{H}=\alpha_r\sqrt{LP}_{R}\textbf{A}\left(\phi\right)\textbf{P}+\dfrac{1}{\sqrt{L}}\nonumber\\
&\times\sum\nolimits_{l=1}^{L}\left(\textbf{G}_{RB}\left(\sum\nolimits_{j=1}^J \mathbf{V}_j^{DL}\mathbf{s}_j^{DL}+\mathbf{c}_0\right)\right.\\
&\left.\qquad+\sum\nolimits_{k=1}^{K}{\mathbf{G}_{RU_{k}}}\left(\mathbf{V}_k^{UL}\mathbf{s}_k^{UL}+\mathbf{c}_{k}^{UL}\right)+\textbf{n}_{R}\right)\mathbf{x}_R^{H}.\nonumber
\end{align}
From~\eqref{Y_hat}, the vectorization of $\hat{\textbf{Y}}$ can be written as
\begin{align}\label{vec_Y_hat}
\hat{\textbf{y}}&=\text{vec}\left(\hat{\textbf{Y}}\right)\nonumber\\
&=\alpha_r\sqrt{LP}_{R}\text{vec}\left(\textbf{A}\left(\phi\right)\textbf{P}\right)\nonumber\\
&\;\;\;+\text{vec}\left(\dfrac{1}{\sqrt{L}}\sum\nolimits_{l=1}^{L}\left(\textbf{G}_{RB}\left(\sum\nolimits_{j=1}^J \mathbf{V}_j^{DL}\mathbf{s}_j^{DL}+\mathbf{c}_0\right)\right.\right.\nonumber\\
&\left.\left.\;\;\;+\sum\nolimits_{k=1}^{K}{\mathbf{G}_{RU_{k}}}\left(\mathbf{V}_k^{UL}\mathbf{s}_k^{UL}+\mathbf{c}_{k}^{UL}\right)+\textbf{n}_{R}\right)\mathbf{x}_R^{H}\right)\nonumber\\
&\triangleq\alpha_r\sqrt{LP}_{R}\text{vec}\left(\textbf{A}\left(\phi\right)\textbf{P}\right)+\boldsymbol\Psi,
\end{align}
where $\boldsymbol\Psi$ is zero-mean, complex Gaussian distributed, and has a non-white covariance matrix of
\begin{align}\label{covariance_matrix_radar}
\boldsymbol\chi= 
 \begin{pmatrix}
\bold{P}^H(  \tilde{\boldsymbol\chi}+ \sigma_{R}^{2}\mathbf{I}_{R})\bold{P} & \cdots & \textbf{0} \\
  \vdots &  \ddots & \vdots \\
  \textbf{0}  & \cdots &\bold{P}^H(  \tilde{\boldsymbol\chi}+ \sigma_{R}^{2}\mathbf{I}_{R})\bold{P}
 \end{pmatrix}.
\end{align}
In the above, $\boldsymbol\chi\!\in\!\mathbb{C}^{R^2\times R^2}$ and $\tilde{\boldsymbol\chi}\!=\!\Big(\!\textbf{G}_{RB}\sum\limits_{j=1}^{J} \left(\!\mathbf{V}_j^{DL}\left(\!\mathbf{V}_j^{DL}\!\right)^H\right.\\ \left.+\psi\text{diag}\left(\!\mathbf{V}_j^{DL} \left(\!\mathbf{V}_j^{DL}\!\right)^H\right)\!\right)\left(\!\mathbf{G}_{RB}\!\right)^H\!\!+\!\!\sum\limits_{k=1}^{K}\!\left(\!{\mathbf{G}_{RU_{k}}}\mathbf{V}_k^{UL}\left(\!\mathbf{V}_k^{UL}\!\right)^{H}\right.\\ \left.\times\left(\mathbf{G}_{RU_{k}}\right)^{H}+\psi\text{diag}\left(\mathbf{V}_k^{UL} \left(\mathbf{V}_k^{UL}\right)^H\right)\right)\Big)$. 
However, the GLRT in \cite{Bekkerman06}  was applied in the presence of white noise only. Hence, to convert the covariance matrix in \eqref{covariance_matrix_radar} into white, we apply a whitening filter $\boldsymbol{\Pi}^H$, obtained after the Cholesky decomposition\footnote{Note that $\boldsymbol\chi$ and $\boldsymbol\chi^{-1}$ are both positive-definite Hermitian matrices.} of $\boldsymbol\chi^{-1}$ as $\boldsymbol\chi^{-1}=\boldsymbol{\Pi}\boldsymbol{\Pi}^H$, with $\boldsymbol{\Pi}$ being a lower triangle matrix.
Now, the hypothesis testing problem in \eqref{hypothesis} can be equivalently rewritten as
\begin{equation}\label{hypothesis_new}
\hat{\textbf{y}} =
    \begin{cases}
      \mathcal{H}_1:\alpha_r\sqrt{LP}_{R}\boldsymbol{\Pi}^H\overline{\textbf{A}}\left(\phi\right)+\boldsymbol{\Pi}^H\boldsymbol\Psi\,,\\
      \mathcal{H}_0:\boldsymbol{\Pi}^H\boldsymbol\Psi\,,
    \end{cases}       
\end{equation}
where $\overline{\textbf{A}}\left(\phi\right)=\text{vec}\left(\textbf{A}\left(\phi\right)\textbf{P}\right)$. If $p\left(\hat{\textbf{y}},\hat{\alpha}_r, \hat{\phi},\mathcal{H}_1\right)$ and $p\left(\hat{\textbf{y}},\mathcal{H}_0\right)$ denote the probability density function under $\mathcal{H}_1$ and $\mathcal{H}_0$, respectively, and $\hat{\alpha}_r$ and $\hat{\phi}$ indicate the ML estimation of ${\alpha_r}$ and ${\phi}$ under hypothesis $\mathcal{H}_1$, which is expressed as $\left[\hat{\alpha}_r,\hat{\phi}\right]=\max_{\alpha_r,\phi}p\left(\hat{\textbf{y}}\mid\hat{\alpha}_r,\hat{\phi},\mathcal{H}_1\right)$, then the GLRT can be given by
\begin{align}\label{hypothesis_detection}
\ln L_{\hat{\textbf{y}}}\left(\hat{\phi}_{ML}\right)=&\dfrac{\left |\overline{\textbf{A}}^{H}\left(\hat{\phi}_{ML}\right)\boldsymbol{\Pi}\boldsymbol{\Pi}^H\hat{\textbf{y}}\right |^{2}}{\left\vert\left\vert\boldsymbol{\Pi}^H\overline{\textbf{A}}^{H}\left(\hat{\phi}_{ML}\right)\right\vert\right\vert^2}\lessgtr\bar{\delta}\,,\\
=&\dfrac{\left |\text{tr}\left(\hat{\textbf{Y}}\mathbf{P}^H\textbf{A}^{H}\left(\hat{\phi}_{ML}\right)\hat{\boldsymbol\chi}^{-1}\right)\right |^{2}}{\text{tr}\left(\textbf{A}\left(\hat{\phi}_{ML}\right)\mathbf{P}\mathbf{P}^H\textbf{A}^{H}\left(\hat{\phi}_{ML}\right)\hat{\boldsymbol\chi}^{-1}\right)}\underset{\mathcal{H}_0}{\overset{\mathcal{H}_1}{\lessgtr}}\bar{\delta}\,,\nonumber
\end{align}
where $\bar{\delta}$ denotes the decision threshold and $\hat{\boldsymbol\chi}=\tilde{\boldsymbol\chi}+ \sigma_{R}^{2}\mathbf{I}_{R}$. According to \cite{Kay98}, the asymptotic statistic  of $L_{\hat{\textbf{y}}}\left(\hat{\phi}_{ML}\right)$ for both the hypothesis is written as
\begin{align}\label{asymptotic_dist}
\ln L_{\hat{\textbf{y}}}\left(\hat{\phi}_{ML}\right)\backsim
    \begin{cases}
      \mathcal{H}_1:\mathcal{X}_{2}^{2}\left(\rho\right)\,,\\
      \mathcal{H}_0:\mathcal{X}_{2}^{2}\,,
    \end{cases}
\end{align}
where $\mathcal{X}_{2}^{2}$ denotes the central chi-squared distributions with two degrees of freedom (DoFs), $\mathcal{X}_{2}^{2}\left(\rho\right)$ is the non-central chi-squared distributions with two DoFs, and $\rho$ indicates the non-central parameter given as
\begin{align}\label{rho_def}
\begin{aligned}
\rho=&\left|\alpha_r\right|^{2}LP_{R}\text{vec}^{H}\left(\textbf{A}\left(\phi\right)\textbf{P}\right)\boldsymbol\chi^{-1}\text{vec}\left(\textbf{A}\left(\phi\right)\textbf{P}\right)\\
=&\Gamma_{R}\sigma_{R}^{2}\text{tr}\left(\mathbf{A}\left(\phi\right)\textbf{P}\textbf{P}^{H}\mathbf{A}^{H}\left(\phi\right)\hat{\boldsymbol\chi}^{-1}\right).
\end{aligned}
\end{align}
where $\Gamma_{R}={\left|\alpha_r\right|^{2}LP_{R}}/{\sigma_{R}^{2}}$. The decision threshold $\bar{\delta}$ is set according to a desired probability of false alarm $P_{FA}$ as
\begin{align}\label{radar_PFA_prob}
\bar{\delta}=&\mathfrak{F}_{\mathcal{X}_{2}^{2}}^{-1}\left(1-P_{FA}\right)\,,
	\end{align}
where $\mathfrak{F}_{\mathcal{X}_{2}^{2}}^{-1}$ denotes the inverse central chi-squared distribution function with two DoFs. The PoD for the MIMO RS can now be given as
\begin{align}\label{radar_detection_prob}
P_{D}=1-\mathfrak{F}_{\mathcal{X}_{2}^{2}(\rho)}\left(\mathfrak{F}_{\mathcal{X}_{2}^{2}}^{-1}\left(1-P_{FA}\right)\right),
	\end{align}
where $\mathfrak{F}_{\mathcal{X}_{2}^{2}(\rho)}$ is the non-central chi-squared distribution function with two DoFs. 

\subsubsection{Projection matrix design at MIMO RS}
In order to mitigate the interference towards the CS, we need to design the projection matrix $\mathbf P$ in such as way that it projects onto the null space of the interference channels $\mathbf W$. Accordingly, to find the null space of  $\mathbf{W}$, we perform its singular value decomposition (SVD), which can be given as $
\mathbf{W}=\mathbf{R}\bold{\Omega}\mathbf{X}^H,
$
where $\mathbf{R}$ and $\mathbf{X}$ are unitary matrices and $\bold{\Omega}$ is a diagonal matrix whose elements are the singular values of $\mathbf{W}$. Now, let 
\begin{align}\label{aaabbb1}
\bold{\bar\Omega}\triangleq\text{diag}(\bar\omega_{1}\dots\bar\omega_{p}),
\end{align} 
where $p\triangleq \min(N_{BS+Users},R_T)$, $\bar\omega_{1}>\bar\omega_{2}>\dots\bar\omega_{q}=\bar\omega_{q+1}=\dots=\bar\omega_{p}=0$ and 
\begin{align}\label{aaabbb2}
\bold{\tilde\Omega}\triangleq\text{diag}(\tilde\omega_{1}\dots\tilde\omega_{R_T}),
\end{align} 
where $\tilde\omega_{r}\triangleq 0, \forall\,r\leq q$, $\tilde\omega_{r}\triangleq 1, \forall \,r> q$, with $\bold{\bar\Omega\tilde\Omega=\bold{0}}$. 
Using the above definitions, the beamforming matrix for NSP can be defined as \cite{Khawar15}
\begin{equation}\label{projection_matrix_eq1}
\mathbf{P}\triangleq\mathbf{X}\bold{\tilde\Omega}\mathbf{X}^H.
\end{equation}
\begin{proposition}\label{NSP_prop}
When ${R}_{T}\gg \left(N_0+JN_j\right)$, the beamforming matrix $\mathbf{P}\in\mathbb{C}^{R_T\times R_T}$ can be  projected orthogonally onto the null-space of the entire CS involving the full set of interference channels $\mathbf{W}\in\mathbb{C}^{N_{BS+Users}\times R_T}$.
\end{proposition}
\begin{IEEEproof}
The proof is given in Appendix \ref{Appendix_A0}.
\end{IEEEproof}

\section{{Spectrum utilising FD MIMO CS}}\label{section_detection_prob}
After establishing the role of the MIMO RS, in this section, we formulate the part played by the FD MIMO CS in the proposed spectrum sharing framework. 
\subsubsection{Achievable rate}
Using the spectrum of  MIMO RS, the signals received at the FD BS and the $j$-th Dl user of the CS at time index $l$ can be written, respectively as
\begin{align}
			\mathbf{y}_0 (l)\!=&\sum\limits_{k=1}^K\mathbf{H}_{k}^{UL}\left(\mathbf{x}_k^{UL}(l)+\mathbf{c}_k^{UL}(l)\right)+\mathbf{H}_{0}\left(\mathbf{x}_0(l)\!+\! \mathbf{c}_0(l)\right)\nonumber\\
		&\qquad+ \mathbf{e}_0(l)+\underbrace{\sqrt{P_R}\mathbf{W}_{BR}^{DL}{\mathbf{s}}_{R}(l)}_{\text{Interference from RS}}+\mathbf{n}_0(l),  \label{i_receive_imperfect3}\\ 
			\mathbf{y}_j^{DL}(l) \!=& \mathbf{H}_{j}^{DL}\left(\mathbf{x}_0(l)\! + \!\mathbf{c}_0(l)\right) \!+\! \sum\limits_{k=1}^K \mathbf{H}_{jk}^{DU}\left(\mathbf{x}_k^{UL}(l)\!+\!\mathbf{c}_k^{UL}(l)\right)\nonumber\\
		&\qquad+ \mathbf{e}_j^{DL}(l)+\underbrace{\sqrt{P_R}\mathbf{W}_{j}^{DL}{\mathbf{s}}_{R}(l)}_{\text{Interference from RS}}+\mathbf{n}_j^{DL}(l),\label{i_receive_imperfect33}
		\end{align} 
where,  $\mathbf{n}_0 (l)\in \mathbb{C}^{N_0}$ and $\mathbf{n}_j^{DL}(l) \in \mathbb{C}^{N_j}$ denote the additive white Gaussian noise (AWGN) vector with zero mean and covariance matrix $\mathbf{R}_0=\sigma_0^2\mathbf{I}_{N_0}$ and $\mathbf{R}_j^{DL}=\sigma_j^2\mathbf{I}_{N_j}$ at the BS and the $j$-th DL user, respectively, and $\mathbf{c}_k^{UL}(l) \sim \mathcal{CN} \left(\mathbf{0}, \psi\: \text{diag}\left(\mathbf{V}_k^{UL}\left(\mathbf{V}_k^{UL}\right)^H\right)\right),~\mathbf{c}_k^{UL}(l) ~ \bot ~\mathbf{x}_k^{UL}(l)$ and $\mathbf{e}_j^{DL} (l) \sim \mathcal{CN} \left(\mathbf{0}, \upsilon \;\text{diag}\left(\mathbf{\Phi}_j^{DL}\right)\right)$, $~\mathbf{e}_j^{DL}(l)$ $~ \bot ~\mathbf{\hat u}_j^{DL}(l)$ are the transmit and receive distortion at the $k$-th UL and  $j$-th DL user, respectively with $\psi \ll 1$ and $\upsilon \ll1$. Here, $\mathbf{\Phi}_j^{DL}= \text{Cov} \{\mathbf{\hat{u}}_j^{DL}(l)\}$ and $\mathbf{\hat{u}}_j^{DL}(l)=\mathbf{y}_j^{DL}(l)-\mathbf{e}_j^{DL}(l)$. The transmitter/receiver distortion $\mathbf{c}_0(l)$/$\mathbf{e}_0(l)$ can be defined similarly. Like before, we drop the time index $l$ henceforth, unless otherwise stated.
  \begin{rem}
Since the MIMO RS transmits NSP based signals, the interference terms in~\eqref{i_receive_imperfect3},~\eqref{i_receive_imperfect33} $\mathbf{W}_{BR}^{DL}{\mathbf{s}}_{R}(l)$ and $\mathbf{W}_{j}^{DL}{\mathbf{s}}_{R}(l)$ will effectively be $\mathbf{W}_{BR}^{DL}{\mathbf{x}}_{R}(l)$ and $\mathbf{W}_{j}^{DL}{\mathbf{x}}_{R}(l)$ and hence, will be mitigated. However, they have been retained here for tractability of the corresponding expressions, but will be ignored in calculation of the numerical results.
\end{rem}   
  \begin{rem}
Since the codeword $\mathbf{x}_0$ and the SI channel $\mathbf{H}_0$ are known to the BS (its own transmitted signal), the term $\mathbf{H}_0\mathbf{x}_0$ in~\eqref{i_receive_imperfect3} can be cancelled out and thus, the remaining part $\mathbf{H}_0\mathbf{c}_0$ can be treated as the RSI. However, the term $\mathbf{H}_0\mathbf{x}_0$ has been retained for tractability, but will be ignored in calculation of the numerical results.
\end{rem}      

\begin{lem}\label{Lemma_1_cov}
		The approximated aggregate interference-plus-noise terms at the $k$-th UL and $j$-th DL user, can be given respectively as\footnote{Note that  approximation of $\boldsymbol{\Sigma}_k^{UL}$ and $\boldsymbol{\Sigma}_j^{DL}$ is a practical assumption~\cite{day2012}. The values of $\psi$ and $\upsilon$ are much lower than 1. However, their values might not be negligible under a strong SI channel~\cite{day2012}.} in~\eqref{interference_covariance_imperfect3} and \eqref{interference_covariance_imperfect4} as shown on the top of the next page.
\begin{figure*}		
		\newcounter{covariance3}
		    \setcounter{covariance3}{\value{equation}}
			\begin{IEEEeqnarray}{rCl} \label{interference_covariance_imperfect3}
				\boldsymbol{\Sigma}_k^{UL} \approx &&\sum\nolimits_{j \neq k}^K \mathbf{H}_j^{UL}  \mathbf{V}_j^{UL} \left( \mathbf{V}_j^{UL}\right)^H \left( \mathbf{H}_j^{UL}\right)^H  + \psi \sum\nolimits_{j =1 }^K \mathbf{H}_j^{UL} \text{diag} \left(  \mathbf{V}_j^{UL} \left( \mathbf{V}_j^{UL}\right)^H\right) \left( \mathbf{H}_j^{UL}\right)^H \nonumber \\
				&& + \sum\nolimits_{j =1}^J\mathbf{H}_0 \left(  \mathbf{V}_j^{DL} \left( \mathbf{V}_j^{DL}\right)^H + \psi \text{diag} \left( \mathbf{V}_j^{DL} \left( \mathbf{V}_j^{DL}\right)^H\right)\right) \mathbf{H}_0^H +P_{R}\left(\mathbf{W}_{BR}^{DL}\left(\mathbf{W}_{BR}^{DL}\right)^{H}\right)+ 
				\sigma_0^2\mathbf{I}_{N_0} \nonumber \\
				&& +\upsilon \sum\nolimits_{j =1}^J  \text{diag} \left( \mathbf{H}_0   \mathbf{V}_j^{DL} \left( \mathbf{V}_j^{DL}\right)^H  \mathbf{H}_0^H\right) +  \upsilon \sum\nolimits_{j =1}^K  \text{diag} \left( \mathbf{H}_j^{UL}   \mathbf{V}_j^{UL} \left( \mathbf{V}_j^{UL}\right)^H \left( \mathbf{H}_j^{UL}\right)^H\right), 
				\\ \label{interference_covariance_imperfect4}
				\boldsymbol{\Sigma}_j^{DL} \approx &&\sum\nolimits_{i \neq j}^J \mathbf{H}_j^{DL} \mathbf{V}_i^{DL} \left( \mathbf{V}_i^{DL}\right)^H\left( \mathbf{H}_j^{DL}\right)^H + \psi \sum\nolimits_{i=1}^J \mathbf{H}_j^{DL} \text{diag} \left( \mathbf{V}_i^{DL} \left( \mathbf{V}_i^{DL}\right)^H\right) \left( \mathbf{H}_j^{DL}\right)^H \nonumber \\
				&& + \sum\nolimits_{k =1}^K\mathbf{H}_{jk}^{DU} \left( \mathbf{V}_k^{UL} \left( \mathbf{V}_k^{UL}\right)^H + \psi \text{diag} \left( \mathbf{V}_k^{UL} \left( \mathbf{V}_k^{UL}\right)^H\right)\right) \left(\mathbf{H}_{jk}^{DU}\right)^H+P_{R}\left(\mathbf{W}_{j}^{DL}\left(\mathbf{W}_{j}^{DL}\right)^{H}\right)+ \sigma_j^2\mathbf{I}_{N_j}  \nonumber \\
				&& +\upsilon \sum\nolimits_{k =1}^K  \text{diag} \left( \mathbf{H}_{jk}^{DU}  \mathbf{V}_k^{UL} \left( \mathbf{V}_k^{UL}\right)^H  \left(\mathbf{H}_{jk}^{DU}\right)^H\right)   +  \upsilon \sum\nolimits_{i =1}^J  \text{diag} \left( \mathbf{H}_j^{DL}  \mathbf{V}_i^{DL} \left( \mathbf{V}_i^{DL}\right)^H \left( \mathbf{H}_j^{DL}\right)^H\right).
			\end{IEEEeqnarray}
			\hrule
			\end{figure*}
			\end{lem}
		\begin{IEEEproof}
		This approximation can be easily proofed from~\eqref{i_receive_imperfect3} and ~\eqref{i_receive_imperfect33} by considering $\psi << 1$ and $\upsilon <<1$ and ignoring the terms $\psi\upsilon$. 
		\end{IEEEproof}

Hence, from~\eqref{i_receive_imperfect3},~\eqref{i_receive_imperfect33} and Lemma \ref{Lemma_1_cov}, the achievable rate for the $k$-th UL user at the BS and the $j$-th DL user can be given as
        \begin{align}
        R_{k}^{UL} =& \log_2\left\vert\textbf{I}_k^{UL}+\left(\mathbf{U}_k^{UL}\right)^{H}\mathbf{H}_{k}^{UL}\mathbf{V}_k^{UL}\left(\mathbf{V}_k^{UL}\right)^{H}\left(\mathbf{H}_{k}^{UL}\right)^{H}\right.\nonumber\\
        &\left.\hspace{4.5em}\times\mathbf{U}_k^{UL}\left(\left(\mathbf{U}_k^{UL}\right)^{H}\boldsymbol{\Sigma}_k^{UL}\mathbf{U}_k^{UL}\right)^{-1}\right\vert\,,\label{sum_rate_k_UL}\\ 
        R_{j}^{DL} =&\log_2\left\vert \textbf{I}_j^{DL}+\left(\mathbf{U}_j^{DL}\right)^{H}\mathbf{H}_{j}^{DL}\mathbf{V}_j^{DL}\left(\mathbf{V}_j^{DL}\right)^{H}\left(\mathbf{H}_{j}^{DL}\right)^{H}\right.\nonumber\\
        &\left.\hspace{4.5em}\times\mathbf{U}_j^{DL}\left(\left(\mathbf{U}_j^{DL}\right)^{H}\boldsymbol{\Sigma}_j^{DL}\mathbf{U}_j^{DL}\right)^{-1}\right\vert.\label{sum_rate_j_DL}
        \end{align}

\subsubsection{Beamformer design at FD MIMO CS}
The main objective of the CS is to provide the cellular users with data by utilising the spectrum of the RS, but without affecting the PoD of RS. Hence, we formulate the beamforming design problem as
\begin{align}\label{opt_p0}
			\textbf{(P0)}\;\;\;\max_{\mathbf{V}}\qquad &   P_{D} \\
			\text{subject to} \left(C.1\right)\;\;\; &    R_{k}^{UL} \geq R_{k,min}^{UL},\qquad\qquad\;\,~k=1,\ldots,K, \nonumber \\ 
			\left(C.2\right)\;\;\; &  R_{k}^{DL} \geq R_{j,min}^{DL},\qquad\qquad\;\,~j=1,\ldots,J, \nonumber \\ 
			\left(C.3\right)\;\;\; &    \text{tr} \left\{\mathbf{V}_{k}^{UL} \left(\mathbf{V}_{k}^{UL}\right)^H\right\} \leq P_k,~k=1,\ldots,K, \nonumber\\ 
			\left(C.4\right)\;\;\; &   \sum\nolimits_{j=1}^{J} \text{tr} \left\{\mathbf{V}_{j}^{DL} \left(\mathbf{V}_{j}^{DL}\right)^H\right\}\leq P_0, \nonumber
			\end{align}  
where $R_{k,min}^{UL}$ in~$\left(C.1\right)$ and $R_{j,min}^{DL}$ in~$\left(C.2\right)$ are the minimum QoS requirements for the $k$-th UL and $j$-th DL users, respectively. Further, the constraints $\left(C.3\right)$ and $\left(C.4\right)$ regulate the transmit powers at the $k$-th UL user and the BS, respectively and $\mathbf{V}=\left\{\mathbf{V}_{k}^{UL},\mathbf{V}_{j}^{DL}\right\}$ denotes the set of all transmit beamforming matrices.
Since $P_{D}$ is a monotonically increasing function with respect to the non-central parameter $\rho$ \cite{Kay98}, we can equivalently reformulate the problem \textbf{(P0)} as
\begin{align}
			\textbf{(P1.A)}\;\;\;&\quad\max_{\mathbf{V}}  \quad  \text{tr}\left(\mathbf{A}\left(\phi\right){\textbf{P}}{\textbf{P}}^{H}\mathbf{A}^{H}\left(\phi\right)\hat{\boldsymbol\chi}^{-1}\right)\label{interf_cst_cog_cell_p1}\\
			&\text{subject to} \quad\left(C.1\right)-\left(C.4\right)\,\,\text{of (\textbf{P0})},\nonumber 
			\end{align}  
			{\begin{lem}\label{Lemma_LB}
A lower bound for $\text{tr}\left(\mathbf{A}\left(\phi\right){\textbf{P}}{\textbf{P}}^{H}\mathbf{A}^{H}\left(\phi\right)\hat{\boldsymbol\chi}^{-1}\right)$ can be given as
\begin{align}
&\text{tr}\left(\mathbf{A}\left(\phi\right){\textbf{P}}{\textbf{P}}^{H}\mathbf{A}^{H}\left(\phi\right)\hat{\boldsymbol\chi}^{-1}\right)\geq\dfrac{\varphi R^2}{I^{RAD}+R\sigma_{R}^{2}},
\end{align}
where  $\varphi=\text{tr}({\textbf{P}}{\textbf{P}}^{H}$) and $I^{RAD}$ is the total interference power from the CS to the MIMO RS, given as
\begin{align} \label{int_power}
			I^{RAD}=& \sum\nolimits_{k=1}^K \text{tr} \left\{  {\mathbf{G}_{RU_{k}}} \left( \mathbf{V}_k^{UL} \left(\mathbf{V}_k^{UL}\right)^H \right.\right. \nonumber\\
			&\hspace{4em} \left.\left. + \psi\text{diag}\left(  \mathbf{V}_k^{UL} \left(\mathbf{V}_k^{UL}\right) ^H\right)\right)({\mathbf{G}_{RU_{k}}})^H\right\} \nonumber \\
			& +  \sum\nolimits_{j=1}^J  \text{tr} \left\{  \mathbf{G}_{RB} \left( \mathbf{V}_j^{DL} \left(\mathbf{V}_j^{DL}\right)^H \right.\right. \\
			&\hspace{4em} \left.\left. +  \psi\text{diag}\left(  \mathbf{V}_j^{DL} \left(\mathbf{V}_j^{DL}\right) ^H\right)\right)(\mathbf{G}_{RB})^H\right\}, \nonumber
		\end{align}
		\end{lem}
	\begin{IEEEproof}
	The proof is given in Appendix \ref{Appendix_A1}.
	\end{IEEEproof}}
{From Lemma~\ref{Lemma_LB}, the problem $\textbf{(P1.A)}$ can be equivalently transformed into a interference minimisation problem as}
\begin{align}
			\textbf{(P1.B)}\;\;\;&\quad\min_{\mathbf{V}}  \qquad I^{RAD} \label{interf_cst_cog_cell_p1}\\
			&\text{subject to} \quad\left(C.1\right)-\left(C.4\right)\,\,\text{of (\textbf{P0})}. \nonumber
			\end{align}  
Now, for analytical simplicity, we write the QoS constraints $\left(C.1\right)$ and $\left(C.2\right)$ in terms of mean squared errors (MSE). To calculate the MSE, we first apply linear receive filters $\mathbf{U}_k^{UL} \in \mathbb{C}^{N_0 \times d_k^{UL}}$ and $\mathbf{U}_j^{DL} \in \mathbb{C}^{N_j \times d_j^{DL}}$ to $\mathbf{y}_0$ and $\mathbf{y}_{j}^{DL}$ to obtain the source symbols. Accordingly, the extracted  $k$-th UL user's symbols at the BS and $j$-th DL user's symbol from the BS can be given by
 \begin{align}
			&\mathbf{\hat{s}}_k^{UL} = \left(\mathbf{U}_k^{UL}\right)^{H}\left(\sum\nolimits_{k=1}^K\mathbf{H}_{k}^{UL}\left( \mathbf{V}_k^{UL}\mathbf{s}_k^{UL}+\mathbf{c}_k^{UL}\right)\right.\label{rcvd_signal_UL_1}\\
			&\left.+\mathbf{H}_{0}\Big(\sum\nolimits_{j=1}^J \mathbf{V}_j^{DL}\mathbf{s}_j^{DL}+ \mathbf{c}_0\Big)+ \mathbf{e}_0+\mathbf{W}_{BR}^{DL}\mathbf{s}_{R}+\mathbf{n}_0\right),\nonumber\\				
			&\mathbf{\hat{s}}_{j}^{DL} = \left(\mathbf{U}_j^{DL}\right)^{H}\left( \mathbf{H}_{j}^{DL}\left(\sum\nolimits_{j=1}^J \mathbf{V}_j^{DL}\mathbf{s}_j^{DL} + \mathbf{c}_0\right) \right.\label{rcvd_signal_DL_1}\\
			&\left.+ \sum\limits_{k=1}^K \mathbf{H}_{jk}^{DU}\left( \mathbf{V}_k^{UL}\mathbf{s}_k^{UL}+\mathbf{c}_k^{UL}\right) + \mathbf{e}_j^{DL}+\mathbf{W}_{j}^{DL}\mathbf{s}_{R}+\mathbf{n}_j^{DL}\right).\nonumber
			\end{align}	    
Hence, the MSE for $k$-th UL and $j$-th DL users can be respectively given as
		\begin{align}
&\mathbf{E}_k^{UL}\left(\{\mathbf{U}\},\{\mathbf{V}\}\right)=\left(\left(\mathbf{U}_k^{UL}\right)^H\mathbf{H}_{k}^{UL}\mathbf{V}_k^{UL}-\mathbf{I}_{d_k^{UL}}\right) \label{mse_matrix_UL}\\
				&\times \left(\left(\mathbf{U}_k^{UL}\right)^H \mathbf{H}_{k}^{UL}\mathbf{V}_k^{UL}-\mathbf{I}_{d_k^{UL}}\right)^H + \left(\mathbf{U}_k^{UL}\right)^H \boldsymbol{\Sigma}_k^{UL} \mathbf{U}_k^{UL},  \nonumber\\
&\mathbf{E}_j^{DL}\left(\{\mathbf{U}\},\{\mathbf{V}\}\right)=\left(\left(\mathbf{U}_j^{DL}\right)^H\mathbf{H}_{j}^{DL}\mathbf{V}_j^{DL}-\mathbf{I}_{d_j^{DL}}\right) \label{mse_matrix_DL}\\
&\times\left(\left(\mathbf{U}_j^{DL}\right)^H \mathbf{H}_{j}^{DL}\mathbf{V}_j^{DL}-\mathbf{I}_{d_j^{DL}}\right)^H  + \left(\mathbf{U}_j^{DL}\right)^H \boldsymbol{\Sigma}_j^{DL} \mathbf{U}_j^{DL},\nonumber
			\end{align}
where $\mathbf{U}=\left\{\mathbf{U}_{k}^{UL},\mathbf{U}_{j}^{DL}\right\}$ denotes the set of all receive beamforming matrices. Note that, for fixed transmit beamforming matrices, the optimal receive beamforming matrix at the BS for the $k$-th UL user and at the $j$-th DL user are MMSE receivers, which can be expressed as~\eqref{rcvd_UL} and \eqref{rcvd_DL} shown on the top of the next page.
\begin{figure*}
 \begin{align}
\mathbf{U}_k^{UL^{\star}}&=\arg\min_{\mathbf{U}_k^{UL}}\text{tr}\left(\mathbf{E}_k^{UL}\right)=\left(\mathbf{V}_k^{UL}\right)^{H}\left(\mathbf{H}_{k}^{UL}\right)^{H}\left(\mathbf{H}_{k}^{UL}\mathbf{V}_k^{UL}\left(\mathbf{V}_k^{UL}\right)^{H}\left(\mathbf{H}_{k}^{UL}\right)^{H} + \boldsymbol{\Sigma}_k^{UL}\right)^{-1}\,;\label{rcvd_UL}\\
\mathbf{U}_j^{DL^\star}&=\arg\min_{\mathbf{U}_j^{DL}}\text{tr}\left(\mathbf{E}_j^{DL}\right)=\left(\mathbf{V}_j^{DL}\right)^{H}\left(\mathbf{H}_{j}^{DL}\right)^{H}\left(\mathbf{H}_{j}^{DL}\mathbf{V}_j^{DL}\left(\mathbf{V}_j^{DL}\right)^{H}\left(\mathbf{H}_{j}^{DL}\right)^{H} + \boldsymbol{\Sigma}_j^{DL}\right)^{-1}\,.\label{rcvd_DL}
	\end{align}	
	\hrule
	\end{figure*}    		
Accordingly, substituting~\eqref{rcvd_UL} and~\eqref{rcvd_DL} in~\eqref{mse_matrix_UL} and~\eqref{mse_matrix_DL} and using the property $\left(\textbf{A} + \textbf{BCD}\right)^{-1} = \textbf{A}^{-1} -\textbf{A}^{-1}\textbf{B}\left(\textbf{D}\textbf{A}^{-1}\textbf{B} + \textbf{C}^{-1}\right)^{-1}\textbf{D}\textbf{A}^{-1}$ \cite{Antonia_2004}, the MSE matrices for the $k$-th UL and $j$-th DL users can be written as	
 \begin{align} 
&\mathbf{E}_{k,\text{MMSE}}^{UL}\left(\{\mathbf{V}\}\right)\label{mse_matrix_UL_theorem}\\
				&=\left(\textbf{I}_k^{UL}+\left(\mathbf{V}_k^{UL}\right)^{H}\left(\mathbf{H}_{k}^{UL}\right)^{H}\left(\boldsymbol{\Sigma}_k^{UL}\right)^{-1}\mathbf{H}_{k}^{UL}\mathbf{V}_k^{UL}\right)^{-1},\nonumber 
				\end{align}
				\begin{align}
&\mathbf{E}_{j,\text{MMSE}}^{DL}\left(\{\mathbf{V}\}\right)\label{mse_matrix_DL_theorem}\\ 
				&=\left(\textbf{I}_j^{DL}+\left(\mathbf{V}_j^{DL}\right)^{H}\left(\mathbf{H}_{j}^{DL}\right)^{H}\left(\boldsymbol{\Sigma}_j^{DL}\right)^{-1}\mathbf{H}_{j}^{DL}\mathbf{V}_j^{DL}\right)^{-1}.\nonumber
	\end{align}	    	
From~\eqref{mse_matrix_UL_theorem}--\eqref{mse_matrix_DL_theorem} and~\eqref{sum_rate_k_UL}--\eqref{sum_rate_j_DL}, we obtain the relation between rate and MSE as
 \begin{align}
        R_{k}^{UL} &= \log_2\left\vert\left(\mathbf{E}_{\text{MMSE},k}^{UL}\left(\{\mathbf{V}\}\right)\right)^{-1}\right\vert\,;\label{sum_rate_mse_UL}\\
        R_{j}^{DL} &=\log_2\left\vert\left(\mathbf{E}_{\text{MMSE},j}^{DL}\left(\{\mathbf{V}\}\right)\right)^{-1}\right\vert\,.\label{sum_rate_mse_Dl}
 \end{align}	     

Before proceeding to the next section, we simplify the notations by combining UL and DL channels similar to~\cite{Biswas16}. Let the symbols $\mathcal{S}^{UL}$ and $\mathcal{S}^{DL}$ denote the set of $K$ UL and $J$ DL channels, respectively, and the channels from RS to BS and the set of $J$ DL channels from RS to DL users are denoted by $\mathcal{S}_{BU}^{DL}$ and $\mathcal{S}_{BR}^{DL}$, respectively. Now, denoting 
		\begin{table}[!t]
	 				\renewcommand{\arraystretch}{1.25}
					\captionsetup{justification=centering}
	 				\caption{Simplification of Notations}
	 				\label{table_simplify1}
	 				\centering
	 				\begin{tabular}{|l|l|}
	 					\hline
	 					$\mathbf{H}_{ij} $ &   $\mathbf{H}_{j}^{UL},   i \in \mathcal{S}^{UL},~ j \in \mathcal{S}^{UL}$; \,\,$\mathbf{H}_{0},  i \in \mathcal{S}^{UL},~ j \in \mathcal{S}^{DL}$;\,\\&\,$\mathbf{H}_{ij}^{DU},  i \in \mathcal{S}^{DL},~ j \in \mathcal{S}^{UL}$;\,\,$\mathbf{H}_{i}^{DL},  i \in \mathcal{S}^{DL},~ j \in \mathcal{S}^{DL}$ \\
	 					\hline
	 					$\mathbf{G}_{R_j} $  & $\mathbf{G}_{RU_j},    j \in \mathcal{S}^{UL}$;\,\,$\mathbf{G}_{RB} ,    j \in \mathcal{S}^{DL}$ \\ 
	 					\hline
	 					$\mathbf{W}_{i} $  & $\mathbf{W}_{i}^{DL},    i \in \mathcal{S}^{DL}$;\,\,$\mathbf{W}_{BR}^{DL} ,    i \in \mathcal{S}_{BR}^{DL}$ \\ 
	 					\hline
	 					$\mathbf{n}_i $  & $\mathbf{n}_0,    i \in \mathcal{S}^{UL}$;\,\,$\mathbf{n}_i^{DL},   i \in \mathcal{S}^{DL}$ \\ 
	 					\hline
	 					$\tilde{N}_{i} (\tilde{M}_{i}) $  & $N_0 \left(M_i\right),    i \in \mathcal{S}^{UL}$;\,\,$N_i \left(M_0\right),    i \in \mathcal{S}^{DL}$ \\ 
	 					\hline
	 				\end{tabular}
	 			\end{table}
		$\mathbf{V}_i^{X}$, $\mathbf{U}_i^{X}$, $d_i^{X}$ and $\boldsymbol{\Sigma}_i^{X}$, $X \in \{UL,DL\}$ as $\mathbf{V}_i$, $\mathbf{U}_i$, $d_i$ and $\boldsymbol{\Sigma}_i$, respectively, and expressing $\mathbf{H}_{ij}$, $\mathbf{G}_{R_j}$, $\mathbf{n}_i$, and receive (transmit) antenna numbers $\tilde{N}_{i} \left(\tilde{M}_{i}\right)$ as shown in Table~\ref{table_simplify1}, the MSE of the $i$-th link, $i \in \mathcal{S} \triangleq \mathcal{S}^{UL} \cup \mathcal{S}^{DL}$ and the interference power at the MIMO RS,  $I^{RAD}$ can be rewritten, respectively as
		  \begin{align} \label{mse_matrix_mixed}
			\mathbf{E}_i 
		= \left(\mathbf{U}_i^H\mathbf{H}_{ii}\mathbf{V}_i-\mathbf{I}_{d_i}\right) \left(\mathbf{U}_i^H \mathbf{H}_{ii}\mathbf{V}_i-\mathbf{I}_{d_i}\right)^H + \:\mathbf{U}_i^H \boldsymbol{\Sigma}_i \mathbf{U}_i,
		\end{align}
where
		  \begin{align}
			\boldsymbol{\Sigma}_i \!=&\! \sum_{j \in \mathcal{S}, j\neq i}\! \mathbf{H}_{ij} \mathbf{V}_j\mathbf{V}_j^H \mathbf{H}_{ij}^H \!+\! \psi \sum_{j \in \mathcal{S}}  \mathbf{H}_{ij} \text{diag}  \left(\mathbf{V}_j\mathbf{V}_j^H\right) \mathbf{H}_{ij}^H\nonumber\\
			&+\! \upsilon \sum\nolimits_{j \in \mathcal{S}}  \text{diag} \left(\mathbf{H}_{ij}   \mathbf{V}_j\mathbf{V}_j^H \mathbf{H}_{ij}^H\right) +P_{R}\left(\textbf{W}_{i}\left(\textbf{W}_{i}\right)^{H}\right)\nonumber\\
			&+ \sigma_i^2\mathbf{I}_{\tilde{N}_i},\label{interference_mixed}
			\end{align}
			\begin{align}
		I^{RAD}\!&=\!\sum\limits_{j \in \mathcal{S}} \text{tr} \left\{\! \mathbf{G}_{R_j}\left( \mathbf{V}_j \mathbf{V}_j^H + \psi\text{diag}\left( \!\mathbf{V}_j \mathbf{V}_j^H\!\right)\!\right)\mathbf{G}_{R_j}^H\!\right\}.  \label{PU_inter}
		\end{align}  
Now, using~\eqref{sum_rate_mse_UL}, \eqref{sum_rate_mse_Dl}, the above simplified notations, and epigraph method, \cite{Boyd04} the optimisation problem \textbf{(P1.B)} can be equivalently reformulated as 
\begin{align}\label{opt_p1C}
			&\textbf{(P1.C)}\;\;\;\min_{\mathbf{V},\Gamma}\qquad  \ \Gamma\\
			&\text{subject to} \quad \left(C.1\right)\;\;\;   \quad {I}^{RAD}\leq \Gamma\,\nonumber\\ 
			&\left(C.2\right)\;\;\;    \log_2\left\vert\left(\mathbf{E}_{i,\text{MMSE}}\left(\{\mathbf{V}_{i}\}\right)\right)^{-1}\right\vert \geq R_{i,min}^{UL},\quad~i\in \mathcal{S}^{UL},\nonumber \\
			&\left(C.3\right)\;\;\;  \log_2\left\vert\left(\mathbf{E}_{i,\text{MMSE}}\left(\{\mathbf{V}_{i}\}\right)\right)^{-1}\right\vert \geq R_{i,min}^{DL},\quad~i\in\mathcal{S}^{DL},  \nonumber\\ 
			&\left(C.4\right)\;\;\;    \text{tr}\left\{\mathbf{V}_i\mathbf{V}_i^H\right\} \leq P_i,\qquad\qquad\qquad\qquad\;\;~i \in \mathcal{S}^{UL},\nonumber \\ 
			&\left(C.5\right)\;\;\;  \sum\nolimits_{i \in \mathcal{S}^{DL}} \text{tr}\left\{\mathbf{V}_i\mathbf{V}_i^H\right\} \leq P_0,\nonumber
			\end{align}  
\section{Robust Beamformer design at FD MIMO CS}\label{robust_beamforming_design}
In order to design a more practical and robust system, in this section, we assume that perfect CSI knowledge of all the concerned channels is unavailable at the CS\footnote{Note that this assumption doesn't obtrude with the assumption that RS has full CSI knowledge of the CS's channels and adheres to Section II.3.}.  By considering the worst-case (norm-bounded error) model~\cite{zhang2012}, the channel uncertainties can be constructed as
		\begin{eqnarray} 
		\mathbf{H}_{ij} \in \mathcal{H}_{ij} &=& \left\{  \tilde{\mathbf{H}}_{ij} + \boldsymbol{\Delta_{ij}} : \lVert \boldsymbol{\Delta_{ij}} \rVert_F  \leq \delta_{ij} \right\}, \label{csi_imperfect}\\
				\mathbf{G}_{R_j} \in \mathcal{G}_{j} &=& \left\{  \tilde{\mathbf{G}}_{R_j} + \boldsymbol{\Lambda} : \lVert \boldsymbol{\Lambda} \rVert_F  \leq \theta,\;\;j\in\mathcal{S} \right\}, \label{csi_imperfect_crn}
		\end{eqnarray}
where the nominal values of the CSI are denoted by $ \tilde{\mathbf{H}}_{ij}$ and  $\tilde{\mathbf{G}}_{j}$, while $\boldsymbol{\Delta_{ij}}$ and $\boldsymbol{\Lambda}_{j}$ represent the channel error matrix and $\delta_{ij}$ and $\theta_{j}$ express the uncertainty bounds.
		\begin{lem}\label{relation_sum_rate_mse}
Assume that $\mathbf{F}\in\mathbb{C}^{d\times d}$ is a positive semi-definite matrix and $\left\vert\mathbf{F}\right\vert\leq 1$. Then the maximisation of the function $\mathit{f}\left(\mathbf{Q}\right)=-\text{tr}\left(\mathbf{Q}\mathbf{F}\right)+\log\left\vert\mathbf{Q}\right\vert+d$ is equivalent to $\log\left\vert\mathbf{F}^{-1}\right\vert$, i.e.,
\begin{align}
\max_{\mathbf{Q}\in\mathbb{C}^{d\times d},\mathbf{Q}\succ 0}\mathit{f}\left(\mathbf{Q}\right)=\log\left\vert\mathbf{F}^{-1}\right\vert\,,
\end{align}  
where the optimum value $\mathbf{Q}^{\text{opt}}=\mathbf{F}^{-1}$.
\end{lem}  
		
Now, applying Lemma~\ref{relation_sum_rate_mse} and decomposing $\textbf{Q}=\mathbf{B}_{i}\mathbf{B}_{i}^{H}$, the problem $\textbf{(P1.C)}$ under channel uncertainties can  be expressed as
		  \begin{align}\label{opt_p2}
			&\textbf{(P2)}\qquad \min_{\mathbf{V},\Gamma}\quad  \Gamma\\
			&\text{subject to}\;\left(C.1\right)  \; I^{RAD} \leq \Gamma,~\forall \mathbf{G}_{R_j} \in \mathcal{G}_{j},\nonumber\\ 
			&\left(C.2\right)  \;\min_{\boldsymbol{\Delta}}\;\max_{\{\mathbf{U}_{i},\textbf{B}_{i}\}}\;-\text{tr}\left(\mathbf{B}_{i}^{H}\mathbf{E}_{i}\mathbf{B}_{i}\right)+2\log\left\vert\mathbf{B}_{i}\right\vert+d_{i} \nonumber\\
			&\hspace{13em}\geq \log(2)R_{i,min}^{UL},\;~i\in\mathcal{S}^{UL},\nonumber \\ 
			&\left(C.3\right)   \;\min_{\boldsymbol{\Delta}}\;\max_{\{\mathbf{U}_{i},\textbf{B}_{i}\}}\;-\text{tr}\left(\mathbf{B}_{i}^{H}\mathbf{E}_{i}\mathbf{B}_{i}\right)+2\log\left\vert\mathbf{B}_{i}\right\vert+d_{i} \nonumber\\
			&\hspace{13em}\geq \log(2)R_{i,min}^{DL},\;~i\in\mathcal{S}^{DL}, \nonumber \\ 
			&\left(C.4\right)  \;  \text{tr}\left\{\mathbf{V}_i\mathbf{V}_i^H\right\} \leq P_i,\qquad\qquad\qquad\;\;~i \in \mathcal{S}^{UL},\nonumber \\ 
			&\left(C.5\right)  \;  \sum\nolimits_{i \in \mathcal{S}^{DL}} \text{tr}\left\{\mathbf{V}_i\mathbf{V}_i^H\right\} \leq P_0, \nonumber
		\end{align}		
		where $\boldsymbol{\Delta}=\{\boldsymbol{\Delta}_{ij}: \forall(i,j)\}$ and $\mathbf{B}_{i}\in\mathbb{C}^{d_{i}\times d_{i}}$, $i\in\mathcal{S}$, is a weight matrix. Due to the constraint $\left(C.1\right)$ in~\eqref{opt_p2} and the inner maximization in $\left(C.2\right)$ and $\left(C.3\right)$, the problem $\textbf{(P2)}$ is intractable. To make the problem tractable, we consider the following max-min inequality  \cite{Jose11} 
		\begin{align}
		\min\limits_{\boldsymbol{\Delta},\boldsymbol{\Lambda}}\;\max\limits_{\{\mathbf{U}_{i},\textbf{Q}_{i}\}}\left(\cdot\right)\geq\max\limits_{\{\mathbf{U}_{i},\textbf{Q}_{i}\}}\;\min\limits_{\boldsymbol{\Delta},\boldsymbol{\Lambda}}\left(\cdot\right) \geq \log(2)R_{i,min}, i\in\mathcal{S}.
		\end{align}
		 Now, we convert $\text{tr}\left(\mathbf{B}_{i}^{H}\mathbf{E}_{i}\mathbf{B}_{i}\right)$ and $I^{RAD}$ into vector forms, given in the following lemma.
		\begin{lem}\label{Lemma_mse_vec_form}
		$\text{tr}\left(\mathbf{B}_{i}^{H}\mathbf{E}_{i}\mathbf{B}_{i}\right)$ and $I^{RAD}$ can be written in the form of vectors as $\text{tr}\left(\mathbf{B}_{i}^{H}\mathbf{E}_{i}\mathbf{B}_{i}\right) =  \left\| \textbf{z}_{ij}  \right\|_2^2$ and $I^{RAD} =\left\|  \boldsymbol{\iota}\right\|_2^2 $, where $\textbf{z}_{ij}$ and $\boldsymbol{\iota}$ are defined as\footnote{For sake of simplicity, we consider $\tilde{M}=M_0=M_i,~i \in \mathcal{S}^{UL} $.}
	\begin{align}\label{mse_vec_form}
	\begin{aligned}
			\textbf{z}_{ij} \!=\!\!& \left[\!\!\!\! \begin{array}{c} 
					\mathrm{vec} \left(\mathbf{B}_{i}^{H}\left(\mathbf{U}_{i}^{H} {\mathbf{H}}_{ii} \mathbf{V}_{i}-\mathbf{I}_{d_i} \right)\right) \\
					\left\lfloor\left( \mathbf{V}_j^T \otimes \left(\mathbf{B}_{i}^{H}\mathbf{U}_i^{H}\right) \right)\mathrm{vec} \left({\mathbf{H}}_{ij}\right)\right\rfloor_{j \in \mathcal{S},j\neq i}   \\
					\left\lfloor \left\lfloor \sqrt{\psi}\left( (\mathbf{\Xi}_{\ell} \mathbf{V}_{j})^T \otimes\left(\mathbf{B}_{i}^{H} \mathbf{U}_{i}^{H}\right) \right) \mathrm{vec} \left({\mathbf{H}}_{ij}\right) \right\rfloor_{\ell \in \mathcal{D}^{({T})}_j} \right \rfloor_{j \in \mathcal{S}}\!\!   \\
					 \left\lfloor \left\lfloor  \sqrt{\upsilon} \left( \mathbf{V}_j^T \otimes \left(\mathbf{B}_{i}^{H}\mathbf{U}_{i}^{H} \mathbf{\Xi}_{\ell}\right)\right) \mathrm{vec} \left({\mathbf{H}}_{ij}\right) \right\rfloor_{\ell \in \mathcal{D}^{({R})}_i}   \right\rfloor_{j \in \mathcal{S}} \!\!  \\
					\!\!\! P_{R}\left(\mathbf{I}_{R}^{T}\otimes \left(\mathbf{B}_{i}^{H}\mathbf{U}_{i}^{H}\right)\right)\mathrm{vec} \left(\textbf{W}_{i}\right)\!\!  \\
					\sigma_i \mathrm{vec} \left(\mathbf{B}_{i}^{H}\mathbf{U}_{i}^{H}\right)
				\end{array}\!\!
				\right]\!\!,\\
				\boldsymbol{\iota}\! =\!\!& \left[\!\! \!
			\begin{array}{c}
				\!	\left\lfloor  \left( \mathbf{V}_j^T \otimes \mathbf{I}_{T} \right)\mathrm{vec} \left(\mathbf{G}_{j}\right) \right\rfloor_{j \in \mathcal{S}}  \\
				\!\sqrt{\psi}\left\lfloor \left\lfloor  \left( (\mathbf{\Xi}_{\ell} \mathbf{V}_{j})^T \otimes \mathbf{I}_{T} \right) \mathrm{vec} \left(\mathbf{G}_{j}\right)  \right \rfloor_{\ell \in \mathcal{D}^{(T)}_j}  \right\rfloor_{j \in \mathcal{S}}  
			\end{array}\!\! \right].
			\end{aligned}
		\end{align}
		Here, $\mathbf{\Xi}_{\ell}$ represents a square matrix with zero as elements, except for the $\ell$-th diagonal element, which is equal to $1$, and $\mathcal{D}^{(R)}_j$ and $\mathcal{D}^{(T)}_j$ denote the set $\{1 \cdots \tilde{N}_j \}$ and $\{1 \cdots \tilde{M}_j \}$, respectively.
	\end{lem}
	\begin{IEEEproof}
	The proof is provided in Appendix \ref{Appendix_A}.
	\end{IEEEproof}
From Lemma \ref{Lemma_mse_vec_form}, the constraint $\left(C.2\right)$ of \textbf{(P2)} can be rewritten as
\begin{align}
 &  \quad\min_{\boldsymbol{\Delta}}\;\max_{\{\mathbf{U}_{i},\textbf{B}_{i}\}}\;\;\; -\text{tr}\left(\mathbf{B}_{i}^{H}\mathbf{E}_{i}\mathbf{B}_{i}\right)+2\log\left\vert\mathbf{B}_{i}\right\vert+d_{i}\nonumber\\
 &\hspace{8em} \geq \log(2)R_{i,min}^{UL},\qquad~i\in\mathcal{S}^{UL},\label{C2_transf_1}\\
&\Rightarrow \begin{cases}\left(C.2a\right)\;\;d_{i}+2\log\left\vert\mathbf{B}_{i}\right\vert-\sum\nolimits_{j\in\mathcal{S}}\lambda_{ij}\\
\hspace{5em}\geq \log(2)R_{i,min}^{UL},\;\forall i\in\mathcal{S}^{UL}\\ \left(C.2b\right)\;\;\max\limits_{\boldsymbol{\Delta}}\parallel\hat{\textbf{z}}_{ij}+\hat{\textbf{Z}}_{ij}\text{vec}\left(\boldsymbol{\Delta}_{ij}\right)\parallel_{2}^{2}\leq \lambda_{ij}\,,\\
\hspace{6em}\parallel\boldsymbol{\Delta}_{ij}\parallel_{F}\leq\delta_{ij}, \forall i,j\in\mathcal{S}^{UL}\,,\end{cases}\label{C2_transf_2}
\end{align}  
where $\hat{\textbf{z}}_{ij}$ and $\hat{\textbf{Z}}_{ij}$ are defined as
		 \begin{align} \label{variables1}
\hat{\textbf{z}}_{ij}\!\!=\!\!& \left[\!\! \begin{array}{c} 
					\mathrm{vec} \left(\mathbf{B}_{i}^{H}\left(\mathbf{U}_{i}^{H} \tilde{\mathbf{H}}_{ii} \mathbf{V}_{i}-\mathbf{I}_{d_i} \right)\right) \\
					\left\lfloor\left( \mathbf{V}_j^T \otimes \left(\mathbf{B}_{i}^{H}\mathbf{U}_i^{H}\right) \right)\mathrm{vec} \left(\tilde{\mathbf{H}}_{ij}\right)\right\rfloor_{j \in \mathcal{S}^{UL},j\neq i}   \\
					\!\!\! \left\lfloor \left\lfloor \sqrt{\psi}\left( (\mathbf{\Xi}_{\ell} \mathbf{V}_{j})^T \otimes\left(\mathbf{B}_{i}^{H} \mathbf{U}_{i}^{H}\right) \right) \mathrm{vec} \left(\tilde{\mathbf{H}}_{ij}\right) \right\rfloor_{\ell \in \mathcal{D}^{({T})}_j} \right \rfloor_{j \in \mathcal{S}}\!\!   \\
					\!\!\! \left\lfloor \left\lfloor  \sqrt{\upsilon} \left( \mathbf{V}_j^T \otimes \left(\mathbf{B}_{i}^{H}\mathbf{U}_{i}^{H} \mathbf{\Xi}_{\ell}\right)\right) \mathrm{vec} \left(\tilde{\mathbf{H}}_{ij}\right) \right\rfloor_{\ell \in \mathcal{D}^{({R})}_i}   \right\rfloor_{j \in \mathcal{S}} \!\!  \\
					\!\!\! P_{R}\left(\mathbf{I}_{R}^{T}\otimes \left(\mathbf{B}_{i}^{H}\mathbf{U}_{i}^{H}\right)\right)\mathrm{vec} \left(\textbf{W}_{i}\right)\!\!  \\
					\sigma_i \mathrm{vec} \left(\mathbf{B}_{i}^{H}\mathbf{U}_{i}^{H}\right)
				\end{array}\!\!
				\right]\,
				\end{align}
				\begin{align}
				\quad \hat{\textbf{Z}}_{i} =&{\left[ 
					\begin{array}{c} 
						\left( \mathbf{V}_i^T \otimes \left(\mathbf{B}_{i}^{H}\mathbf{U}_{i}^{H}\right) \right) \\
						\left\lfloor\left( \mathbf{V}_j^T \otimes \left(\mathbf{B}_{i}^{H}\mathbf{U}_{i}^{H}\right) \right) \right\rfloor_{j \in \mathcal{S}^{UL},j\neq i}   \\
						\!\! \left\lfloor \left\lfloor \sqrt{\psi}\left( (\mathbf{\Xi}_{\ell} \mathbf{V}_{j})^T \otimes \left(\mathbf{B}_{i}^{H}\mathbf{U}_{i}^{H}\right)\right)  \right\rfloor_{\ell \in \mathcal{D}^{({T})}_{j}} \right \rfloor_{j \in \mathcal{S}}  \\
						\!\! \left\lfloor \left\lfloor  \sqrt{\upsilon} \left( \mathbf{V}_{j}^{T} \otimes \left(\mathbf{B}_{i}^{H}\mathbf{U}_{i}^{H} \mathbf{\Xi}_{\ell} \right) \right)\right\rfloor_{\ell \in \mathcal{D}^{({R})}_i}   \right\rfloor_{j \in \mathcal{S}}   \\
						\mathbf{0}_{d_i \tilde{N}_i \times \tilde{N}_i\tilde{M}}						\\
						\mathbf{0}_{d_i \tilde{N}_i \times \tilde{N}_i\tilde{M}}
					\end{array} \right]}.
			\end{align}  
Now, using Lemma \ref{lemma5} from Appendix \ref{Appendix_B}, we relax the semi-infiniteness of the constraint $\left(C.2b\right)$ in~\eqref{C2_transf_2} and then apply Lemma \ref{lemma6} from Appendix \ref{Appendix_B}, to convert the norm constraint of $\left(C.2b\right)$ into a LMI form as			 
		\begin{align}
		\left(C.2b\right)\;\;&		\left[
		\begin{array}{cc} 
		\lambda_{ij}	&\hat{\textbf{z}}_{ij}^{H} \\
		\hat{\textbf{z}}_{ij} & \mathbf{I}_{d_{i}d_{j}}
		\end{array} \right]\nonumber\\
		&\hspace{2.5em}
		+ 
		\left[
		\begin{array}{cc} 
		0	&\text{vec}\left(\boldsymbol{\Delta}_{ij}\right)^{H}\hat{\textbf{Z}}_{ij}^{H} \\
		\hat{\textbf{Z}}_{ij}\text{vec}\left(\boldsymbol{\Delta}_{ij}\right) & \mathbf{0}_{d_{i}d_{j} \times d_{i}d_{j}}
		\end{array} \right]\succeq 0\,.
		\end{align}  
		By choosing
			 \begin{align}
			\mathbf{A} &=\left[
		\begin{array}{cc} 
		\lambda_{ij}	&\hat{\textbf{z}}_{ij}^{H} \\
		\hat{\textbf{z}}_{ij} & \mathbf{I}_{d_{i}d_{j}}
		\end{array} \right],~ \mathbf{P} = \left[ \mathbf{0}_{\tilde{N}_i\tilde{M} \times 1},~ \hat{\textbf{Z}}_{ij}^{H}\right], \nonumber\\
			\mathbf{X} &= \mathrm{vec}\left(\boldsymbol{\Delta_{ij}}\right),\qquad\;~\mathbf{Q} = \left[-1, \mathbf{0}_{1 \times d_{ij}}\right],&
			\end{align}    
			and applying Lemma~\ref{lemma5}, the constraint $\left(C.2b\right)$ of~\eqref{C2_transf_2} is equivalently rewritten as 
			 \begin{align} \label{LMI_c2b}
			\left[\!\!
			\begin{array}{ccc} 
			\lambda_{ij}-\epsilon_{ij}	&\hat{\textbf{z}}_{ij}^{H} & \mathbf{0}_{1 \times \tilde{N}_i\tilde{M}} \\
			\hat{\textbf{z}}_{ij} & \mathbf{I}_{d_{i}d_{j}} & -\delta_{ij} \hat{\textbf{Z}}_{ij}\\
			\mathbf{0}_{\tilde{N}_i\tilde{M} \times 1} & -\delta_{ij}\hat{\textbf{Z}}_{ij}^{H} & \epsilon_{ij} \mathbf{I}_{\tilde{N}_i\tilde{M}}
			\end{array} \!\!
			\right] &\!\!\succeq\!\!& 0,~\forall i,j \in \mathcal{S}^{UL}, \\
			\epsilon_{ij} &\!\!\geq\!\!& 0,~\forall i,j \in \mathcal{S}^{UL},
			\end{align}    
where $\boldsymbol{\epsilon}=\{\epsilon_{ij}:\forall (i,j)\}$. Similar to the transformation of the constraint $\left(C.2\right)$ of \textbf{(P2)}, the constraint $\left(C.3\right)$ of \textbf{(P2)} can be equivalently rewritten as			 
\begin{align}
&  \quad\min_{\boldsymbol{\Delta}}\;\max_{\{\mathbf{U}_{i},\textbf{B}_{i}\}}\;\;\;   -\text{tr}\left(\mathbf{B}_{i}^{H}\mathbf{E}_{i}\mathbf{B}_{i}\right)+2\log\left\vert\mathbf{B}_{i}\right\vert+d_{i} \nonumber\\
&\hspace{8em}\geq \log(2)R_{i,min}^{DL},\qquad~i\in\mathcal{S}^{DL},\label{C2_transf_1}\\
&\Rightarrow \begin{cases}\left(C.3a\right)\;\;d_{i}+2\log\left\vert\mathbf{B}_{i}\right\vert-\sum\nolimits_{j\in\mathcal{S}}\lambda_{ij}\\
\hspace{5em}\geq \log(2)R_{i,min}^{DL},\;\forall i\in\mathcal{S}^{DL}\\ \left(C.3b\right)\;\;\max\limits_{\boldsymbol{\Delta}}\parallel\hat{\textbf{z}}_{ij}+\hat{\textbf{Z}}_{ij}\text{vec}\left(\boldsymbol{\Delta}_{ij}\right)\parallel_{2}^{2}\leq \lambda_{ij}\,,\\
\hspace{5em}\parallel\boldsymbol{\Delta}_{ij}\parallel_{F}\leq\delta_{ij}, \forall i,j\in\mathcal{S}^{DL}\,,\end{cases}\label{C2_transf_2}
\end{align}  
where $\left(C.3b\right)$ is expressed as
			 \begin{eqnarray} \label{LMI_c3b}
			\!\!\!\!\!\!\!\!\!\!\left[\!\!\!\!
			\begin{array}{ccc} 
			\lambda_{ij}-\epsilon_{ij}	&\hat{\textbf{z}}_{ij}^{H} & \mathbf{0}_{1 \times \tilde{N}_i\tilde{M}} \\
			\hat{\textbf{z}}_{ij} & \mathbf{I}_{d_{i}d_{j}} & -\delta_{ij} \hat{\textbf{Z}}_{ij}\\
			\mathbf{0}_{\tilde{N}_i\tilde{M} \times 1} & -\delta_{ij}\hat{\textbf{Z}}_{ij}^{H} & \epsilon_{ij} \mathbf{I}_{\tilde{N}_i\tilde{M}}
			\end{array} \!\!
			\right] \!\!\!\!\!\!&\succeq&\!\! \!\!\!\!0,~\forall i\in \mathcal{S}^{DL},j \in \mathcal{S}, \\
			\epsilon_{ij}\!\!\!\!\!\! &\geq&\!\!\!\!\!\! 0,~\forall i\in \mathcal{S}^{DL},j \in \mathcal{S}.
			\end{eqnarray}    
In similar way, we also transform the constraint $\left(C.1\right)$ of \textbf{(P2)} as
			 \begin{eqnarray}  
			\left[
			\begin{array}{cc} 
			\Gamma	& \boldsymbol{\tilde{\iota}}^H \\
			\boldsymbol{\tilde{\iota}} & \mathbf{I}_{B}
			\end{array} \right]
			+ 
			\left[
			\begin{array}{cc} 
			0	&\boldsymbol{\iota}_{\Lambda}^H \\
			\boldsymbol{\iota}_{\Lambda} & \mathbf{0}_{B \times B}
			\end{array} \right]\succeq 0,
			\end{eqnarray}    
			where $B=R\sum\nolimits_{j\in\mathcal{S}}\left(d_{j}+\tilde{M}_{j}\right)$,
			  \begin{align} \label{variables3}
					\boldsymbol{\tilde{\iota}} &= \left[
					\begin{array}{c}
						\left\lfloor  \left(\mathbf{V}_j^T \otimes \mathbf{I}_{T} \right)\mathrm{vec} \left(\tilde{\mathbf{G}}_{lj}\right) \right\rfloor_{j \in \mathcal{S}}  \\
						\sqrt{\psi}\left\lfloor \left\lfloor \left( (\mathbf{\Xi}_{\ell} \mathbf{V}_{j})^T \otimes \mathbf{I}_{T}\right) \mathrm{vec} \left(\tilde{\mathbf{G}}_{lj}\right)  \right \rfloor_{\ell \in \mathcal{D}^{({T})}_j} \right\rfloor_{j \in \mathcal{S}}  
					\end{array} \right],\nonumber\\
					 \boldsymbol{\iota}_{\Lambda} &= \underbrace{\left[\!\!
						\begin{array}{c}
								\left\lfloor  \left( \mathbf{V}_j^T \otimes \mathbf{I}_{R} \right) \right\rfloor_{j \in \mathcal{S}}  \\
							\sqrt{\psi}\left\lfloor \left\lfloor \left( (\mathbf{\Xi}_{\ell} \mathbf{V}_{j})^T \otimes\mathbf{I}_{R} \right)  \right \rfloor_{\ell \in \mathcal{D}^{({T})}_j}  \right\rfloor_{j \in \mathcal{S}}  
						\end{array} \hspace{-0.20em}\right]}_{\mathbf{E}_{\Lambda}}\!\!\mathrm{vec} \left(\boldsymbol{\Lambda}\right).
				\end{align}    	
				Finally, the constraint $\left(C.1\right)$ of \textbf{(P2)} is equivalently rewritten as
				 \begin{eqnarray} \label{LMI_inter}
					\!\!	\!\!\!\left[
					\begin{array}{ccc} 
						\Gamma-\eta	&\boldsymbol{\tilde{\iota}}^H & \mathbf{0}_{1\times R \tilde{M}} \\
						\boldsymbol{\tilde{\iota}} & \mathbf{I}_{B} & -\theta \mathbf{E}_{\Lambda}\\
						\mathbf{0}_{R \tilde{M} \times 1} & -\theta \mathbf{E}_{\Lambda}^H & \eta \mathbf{I}_{R \tilde{M}}
					\end{array} 
					\right] &\succeq& 0, \\
					\!\!\!\!\!	\eta &\geq& 0.
				\end{eqnarray}    
Using the relaxed LMIs in~\eqref{LMI_c2b}, ~\eqref{LMI_c3b} and~\eqref{LMI_inter}, the problem~\textbf{(P2)} can be written as a SDP problem, given as	 	  
\begin{IEEEeqnarray}{rCl}\IEEEyesnumber \label{opt_P3}  
		&& \textbf{(P3)}\;\;\; \; \min_{\mathbf{V},\Gamma,\mathbf{U},\mathbf{B}, \boldsymbol{\lambda}, \boldsymbol{\epsilon} \geq 0,\eta \geq 0 }  \Gamma\IEEEyesnumber  \\ 
		&&\text{subject to}	\;	\left(C.1\right)
	     \left[
					\begin{array}{ccc} 
						\Gamma-\eta	&\boldsymbol{\tilde{\iota}}^H & \mathbf{0}_{1\times R \tilde{M}} \\
						\boldsymbol{\tilde{\iota}} & \mathbf{I}_{B} & -\theta \mathbf{E}_{\Lambda}\\
						\mathbf{0}_{R \tilde{M} \times 1} & -\theta \mathbf{E}_{\Lambda}^H & \eta \mathbf{I}_{R \tilde{M}}
					\end{array} 
					\right] \succeq 0, \IEEEyesnumber \nonumber \\ 
		&& \left(C.2a\right) d_{i}+2\log\left\vert\mathbf{B}_{i}\right\vert\!-\!\sum\limits_{j\in\mathcal{S}}\lambda_{ij}\geq \log(2)R_{i,min}^{UL},\;\forall i\in\mathcal{S}^{UL},\nonumber  \\
		&& \left(C.2b\right)
			\left[
			\begin{array}{ccc} 
			\lambda_{ij}-\epsilon_{ij}	&\hat{\textbf{z}}_{ij}^{H} & \mathbf{0}_{1 \times \tilde{N}_i\tilde{M}} \\
			\hat{\textbf{z}}_{ij} & \mathbf{I}_{d_{i}d_{j}} & -\delta_{ij} \hat{\textbf{Z}}_{ij}\\
			\mathbf{0}_{\tilde{N}_i\tilde{M} \times 1} & -\delta_{ij}\hat{\textbf{Z}}_{ij}^{H} & \epsilon_{ij} \mathbf{I}_{\tilde{N}_i\tilde{M}}
			\end{array} 
			\right] \succeq 0,\nonumber\\
			&&\hspace{15em}~\forall i\in \mathcal{S}^{UL} ,j \in \mathcal{S}, \nonumber\\
		&& \left(C.3a\right)  d_{i}+2\log\left\vert\mathbf{B}_{i}\right\vert\!-\!\sum\limits_{j\in\mathcal{S}}\lambda_{ij}\geq \log(2)R_{i,min}^{DL},\;\forall i\in\mathcal{S}^{DL}, \IEEEyesnumber \nonumber \\
		&& \left(C.3b\right) \left[
			\begin{array}{ccc} 
			\lambda_{ij}-\epsilon_{ij}	&\hat{\textbf{z}}_{ij}^{H} & \mathbf{0}_{1 \times \tilde{N}_i\tilde{M}} \\
			\hat{\textbf{z}}_{ij} & \mathbf{I}_{d_{i}d_{j}} & -\delta_{ij} \hat{\textbf{Z}}_{ij}\\
			\mathbf{0}_{\tilde{N}_i\tilde{M} \times 1} & -\delta_{ij}\hat{\textbf{Z}}_{ij}^{H} & \epsilon_{ij} \mathbf{I}_{\tilde{N}_i\tilde{M}}
			\end{array} 
			\right] \succeq 0,\nonumber\\
			&&\hspace{15em}~\forall i \in \mathcal{S}^{DL} ,j \in \mathcal{S}, \IEEEyesnumber \nonumber \\ 
		&& \left(C.4\right) \lVert \mathrm{vec} \left(\mathbf{V}_i\right) \rVert_2^2 \leq P_i,~i \in \mathcal{S}^{UL}, \IEEEyesnumber \nonumber \\
		&& \left(C.5\right)\lVert \left\lfloor \mathrm{vec} \left(\mathbf{V}_i\right) \right\rfloor_{i \in \mathcal{S}^{DL}} \rVert_2^2 \leq P_0, \IEEEyesnumber \nonumber 
	\end{IEEEeqnarray}
where $\mathbf{B}=\{\textbf{B}_{i},\;\forall i\in\mathcal{S}\}$ and $\boldsymbol{\lambda}=\{\lambda_{ij},\;\forall i,j\in\mathcal{S}\}$. 

Note that the optimisation problem~\textbf{(P3)} is not jointly convex over the optimisation variables $\mathbf{V}$,  $\mathbf{U}$ and $\mathbf{B}$. However, it is separately convex over each of the variables. Therefore, we adopt an alternating algorithm to solve the problem. This alternating minimisation process is continued until a stationary point is obtained, or a pre-defined number of iterations is reached.  In the following section, we provide details on the spectrum sharing algorithm, including the alternating optimisation of the above SDP problem.

\section{{Spectrum Sharing Algorithm}}
In this section we summarise the roles played in spectrum sharing by the RS and CS  in the form of algorithms. While Algorithm~1 presents the role played by the RS, Algorithm~2 illustrates the role of the CS in the proposed 2 tier spectrum sharing framework\footnote{Note that both algorithms are processed within the same coherence time interval.}. In particular, Algorithm~1 performs  NSP towards the interference channels of the CS, thereby cancelling the interference from RS to CS. Alternatively, Algorithm~2 produces the optimal beamforming matrices at the transmitters and receivers of the CS, to minimise the interference from the CS to RS (equivalently maximises the PoD of RS), while maintaining a particular QoS for the cellular users.  Note that Algorithm~2 is iterative in nature and solves a SDP problem in each iteration, which makes it computationally intensive. Below we provide some qualitative analysis on the complexity of Algorithm~2.
\begin{table}[t!]
\centering
\begin{tabular}{|l|}
\hline
\rowcolor{LightGray}
 {\textbf{{Algorithm 1: Spectrum Sharing Phase at RS}}}  \\
\hline
\rowcolor{Gray}
{\textbf{I. Phase 1 [Initial Phase]:}} \\
A: Obtain CSI of $\{\mathbf{W}_{i}^{DL},   \forall i \in \mathcal{S}^{DL}$,\,\,$\mathbf{W}_{BR}^{DL}\}\subseteq \mathbf W$ through feedback\\ \quad\;\,from CS.\\
\rowcolor{Gray}
{\textbf{II. Phase 2 [Null-space Projection Phase]:}}\\
A: Perform SVD of $\mathbf W$, which is obtained from Phase 1.\\
B: Construct $ \bold{\bar\Omega}$ and $ \bold{\tilde\Omega}$ according to \eqref{aaabbb1} and \eqref{aaabbb2}\\
C: Design the projection matrix $\mathbf P$ based on \eqref{projection_matrix_eq1}.\\
C:  \textbf{Output:} Perform NSP by transmitting waveform $\mathbf{x}_R={\mathbf{P}}\mathbf{s}_{R}$.\\
\hline
\end{tabular}
\end{table}
\begin{table}[t!]
\centering
\begin{tabular}{|l|}
\hline
\rowcolor{LightGray}
 {\textbf{{Algorithm 2: Spectrum Sharing Phase at CS}}}  \\
\hline
\rowcolor{Gray}
{\textbf{I. Phase 1 [Initial Phase]:}} \\
A: Obtain partial CSI of $\{\mathbf{G}_{RU_j}, \forall   j \in \mathcal{S}^{UL}$,\,\,$\mathbf{G}_{RB}\}$.\\
B: Set minimum QoS requirements for UL and DL users: $R_{i,min}^{UL}$,\\ \quad\;\,and $R_{i,min}^{DL}$.\\
C.  Initialize $\mathbf{V}^{[n]}$, $\mathbf{U}^{[n]}$ and $\mathbf{B}^{[n]}$.\\
D. Set iteration number $n=0$, maximum iteration number $=n_{max}$.\\
\rowcolor{Gray}
{\textbf{II. Phase 2 [Beamforming Design Phase (Alternating Approach)]:}}\\
A:  $n \leftarrow n+1$. For fixed $\mathbf{U}_i^{[n-1]}$ and $\mathbf{B}_i^{[n-1]}$, update $\mathbf{V}^{[n]},~\forall i \in \mathcal{S}$\\ \quad\;\,by solving problem $(\textbf{P3})$.\\
B: Update ${\textbf{B}}_{i}^{[n]},~i \in \mathcal{S}$ by solving the problem~\textbf{(P3)} for fixed $\mathbf{U}_i^{[n-1]}$\\ \quad\;\,and $\mathbf{V}_i^{[n-1]}$.\\
C: Update $\mathbf{U}_i^{[n]},~i \in \mathcal{S}$ by solving the problem~\textbf{(P3)} for fixed $\mathbf{V}_i^{[n-1]}$\\ \quad\;\,and $\mathbf{B}_i^{[n-1]}$.\\
D: Repeat steps II.A -- II.C until convergence or $n=n_{max}$.\\
E: \textbf{Output:} Optimal transceivers: $\{\mathbf{U}^{\star},\mathbf{V}^{\star}\}$.\\
\hline
\end{tabular}
\end{table}
\subsubsection*{Computational complexity of Algorithm 2}\label{Comp_complexity}
The computational complexity mainly depends on the number of arithmetic operations required to process Phase 2 of Algorithm~2. In particular, a SDP problem is solved in Phase~2 in three steps, i.e., Step II.A to Step II.C.
Hence, for comparison we first consider a standard real-valued SDP problem as
  \begin{eqnarray}  \label{eq:both5} 
 \min_{\mathbf{x} \in \mathcal{R}^n} \quad && \quad  \mathbf{c}^T\mathbf{x}  \\ 
 \text{subject to\:} && \quad  \mathbf{A}_0 + \sum\nolimits_{i=1}^{n}x_i \mathbf{A}_i \succeq \mathbf{0},~{\text{and}}\quad  \lVert \mathbf{x}\rVert_2 \leq X, \nonumber
 \end{eqnarray}    
where $\mathbf{A}_i$ is a symmetric block-diagonal matrix. If $P$ is the diagonal block of matrix $\mathbf{A}_i$ of size $a_l \times a_l,~l=1, \ldots,P$, then the number of arithmetic operations required to solve \eqref{eq:both5} is upper-bounded by~\cite{bental}
	  \begin{eqnarray} \label{comp_comp}
	 	\mathcal{O}\left(1\right)\left(1+\sum\limits_{l=1}^{P}a_l\right)^{1/2}n\left(n^2 + n 
	 	\sum\limits_{l=1}^P a_l^2 + \sum\limits_{l=1}^{P}a_l^3\right).
	 \end{eqnarray}    
 Thus, using~\eqref{comp_comp}, we can compute the total number of  arithmetic operations required to find the optimal  $\mathbf{V}_i$, $\mathbf{U}_i$, and $\mathbf{B}_i$ in Algorithm~2. For instance, in order to find $\mathbf{V}_i$, the number of diagonal blocks $P$ is $\left|S^{UL}\right| (\left|S\right| + 1) + \left|S^{DL}\right|\left|S\right| + 1$. The constraints~$(C.2b)$ and~$(C.3b)$ create the blocks of size $a_{ij}=2(N_{i}\tilde{M} + d_{i}d_{j}+1)\,,~i \in \mathcal{S}^{DL},~j\in\mathcal{S}^{UL}$ and $a_{ij}=2(\tilde{N}_{i}\tilde{M} + d_{i}d_{j}+1)\,,~i \in \mathcal{S}^{UL},~j\in\mathcal{S}^{DL}$, respectively. The  size of blocks due to the constraint $(C.1)$ is $a = 2(B + R \tilde{M}+1)$, while the constraints for UL power in $(C.4)$ and the BS power in $(C.5)$ make the blocks of size $a_i = \tilde{M} d_i^{UL}+1,~i \in \mathcal{S}^{UL}$ and $a_i=\tilde{M} \sum_{i \in \mathcal{S}^{DL}}d_i^{DL}+1$, respectively. Furthermore, the number required to compute the unknown variables is $n=\sum_{i \in \mathcal{S}} 2\tilde{M} d_i + 2\lvert \mathcal{S}\rvert+2$, where the term $\sum_{i \in \mathcal{S}} 2\tilde{M} d_i$ correlates with the real and image parts of $\mathbf{V}_i$ and the remaining terms are due to the additional slack variables. Similarly, the number of arithmetic operations required for $\mathbf{U}_i,i \in \mathcal{S}$ and $\mathbf{B}_i,i \in \mathcal{S}$ can be calculated.

\section{Numerical Results}\label{section_simulation_results}
\begin{figure*}[t!]
\centering
\begin{minipage}{.5\textwidth}
  \centering
  \includegraphics[width=1\linewidth]{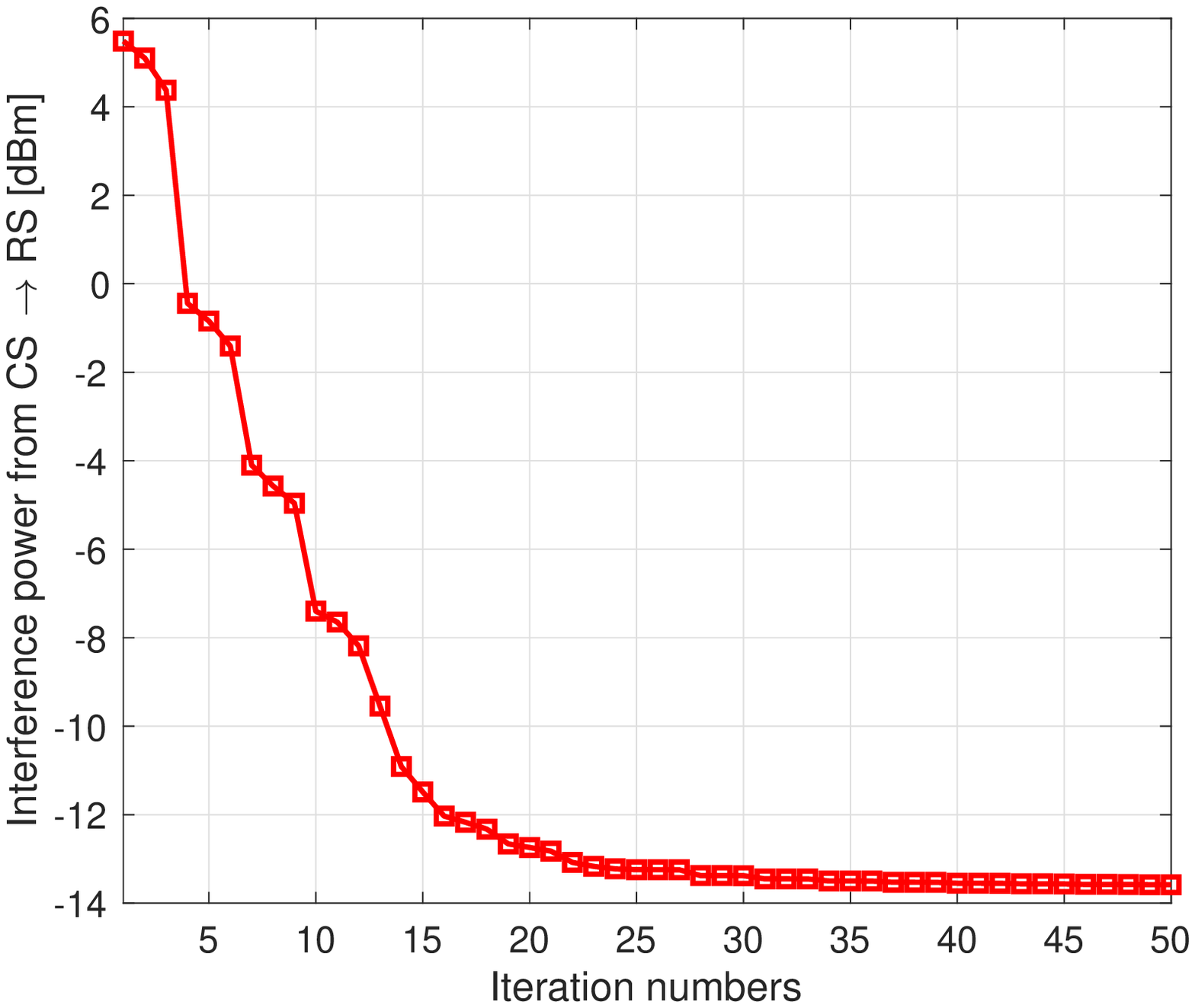}
  \captionof{figure}{Convergence of Algorithm 2.}
  \label{sim_fig_1}
\end{minipage}%
\begin{minipage}{.5\textwidth}
  \centering
  \includegraphics[width=1\linewidth]{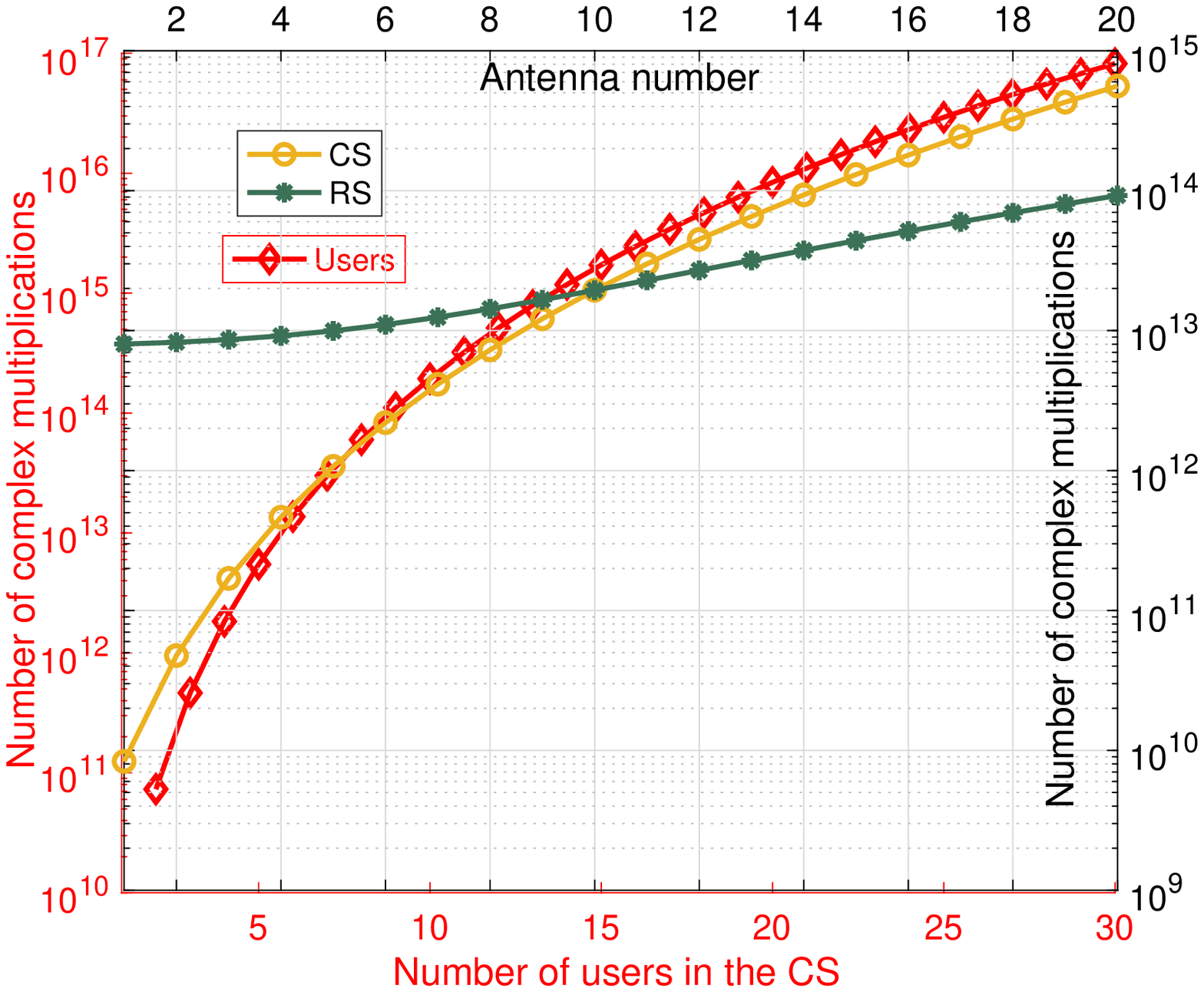}
  \captionof{figure}{{Complexity of Algorithm 2.}}
  \label{sim_fig_2}
\end{minipage}
\end{figure*}
In this section, the performance of the proposed two-tier coexistence framework between a FD MU MIMO CS and a MIMO RS is analysed  with the help of computer simulations\footnote{For reference, the numerical results are obtained using MATLAB R2016b on a Linux server with Intel Xeon processor (16 cores, each clocked at 2 GHz) having 31.4 GiB of memory.} under consideration of QoS of cellular users. The maximum number of iterations is set as $50$ with a tolerance value of $10^{-4}$. The initialisation points are selected using right singular matrices initialisation~\cite{chen2012} and the results are averaged over 100 independent channel realisations.
\subsubsection{Simulation Setup}
To model the CS, we consider small cell deployments under the 3GPP LTE specifications. The motivation behind this are: 1) due to low transmit powers, short transmission distances and low mobility, small cells are considered suitable for implementation of FD technology \cite{Cirik_tvt_2017}, and 2) FCC has proposed the use of small cells in the 3.5 GHz band  for spectrum sharing \cite{FCC12}. Accordingly, a single hexagonal cell of radius $r=40$ m is considered, where the FD BS is located at the centre of the cell and a MIMO RS is located 400 m away from the circumference of the cell. The number of UL and DL users is set as $K=J=2$ and each user, equipped with $N$ antennas is randomly located in the cell. For simplicity, we consider $M_0=N_0=N=\tilde N$.
Next, to model the path loss in the CS, we consider the close-in (CI) free space reference distance path loss model as given in \cite{Andersen_1995}. The CI model is a generic model that describes the large-scale propagation path loss at all relevant frequencies ($> 2$ GHz). This model can be easily implemented in existing 3GPP models by replacing a floating constant with a frequency-dependent constant that represents free space path loss in the first meter of propagation and is given as $PL(f,d)=PL_F(f,d_0)+10\alpha_c\log_{10}\left({d}/{d_0}\right)+\mathcal{X}_{\sigma},\, d>d_0$. Here, $d_0$  is a reference distance at which or closer to, the path loss inherits the characteristics of free-space path loss $PL_F$. Further, $f$ is the carrier frequency, $\alpha_c$ is the path loss exponent, $d$ is the distance between the transmitter and receiver and $\mathcal{X}_{\sigma}$ is the shadow fading standard deviation. We consider $d_0=1$ m, $B=100$ MHz, and carrier frequency $=3.6$ GHz{\footnote{The framework presented in this paper is not limited to any particular frequency band and can also be utilized in other spectrums proposed for sharing around the world, such as 2-4 GHz in the UK, 2.3-2.4 GHz in Europe, etc., albeit certain changes in frequency dependent path loss, line of sight propagation parameters, etc.}}.
 \begin{figure*}[t!]
\centering
\begin{minipage}{.5\textwidth}
  \centering
  \includegraphics[width=1\linewidth]{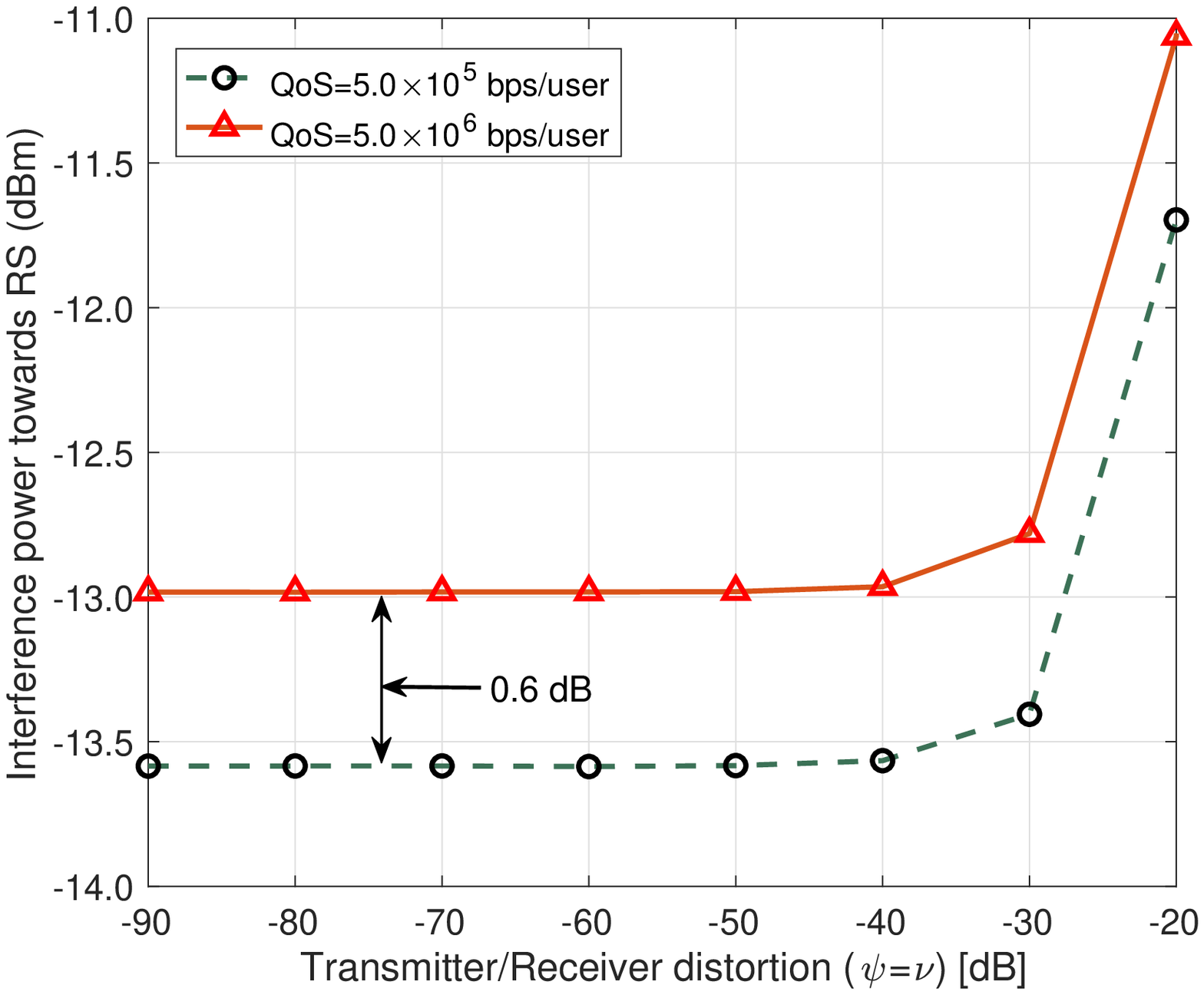}
  \captionof{figure}{Interference power towards RS vs. RSI at CS.}
  \label{sim_fig_5}
\end{minipage}%
\begin{minipage}{.5\textwidth}
  \centering
  \includegraphics[width=1\linewidth]{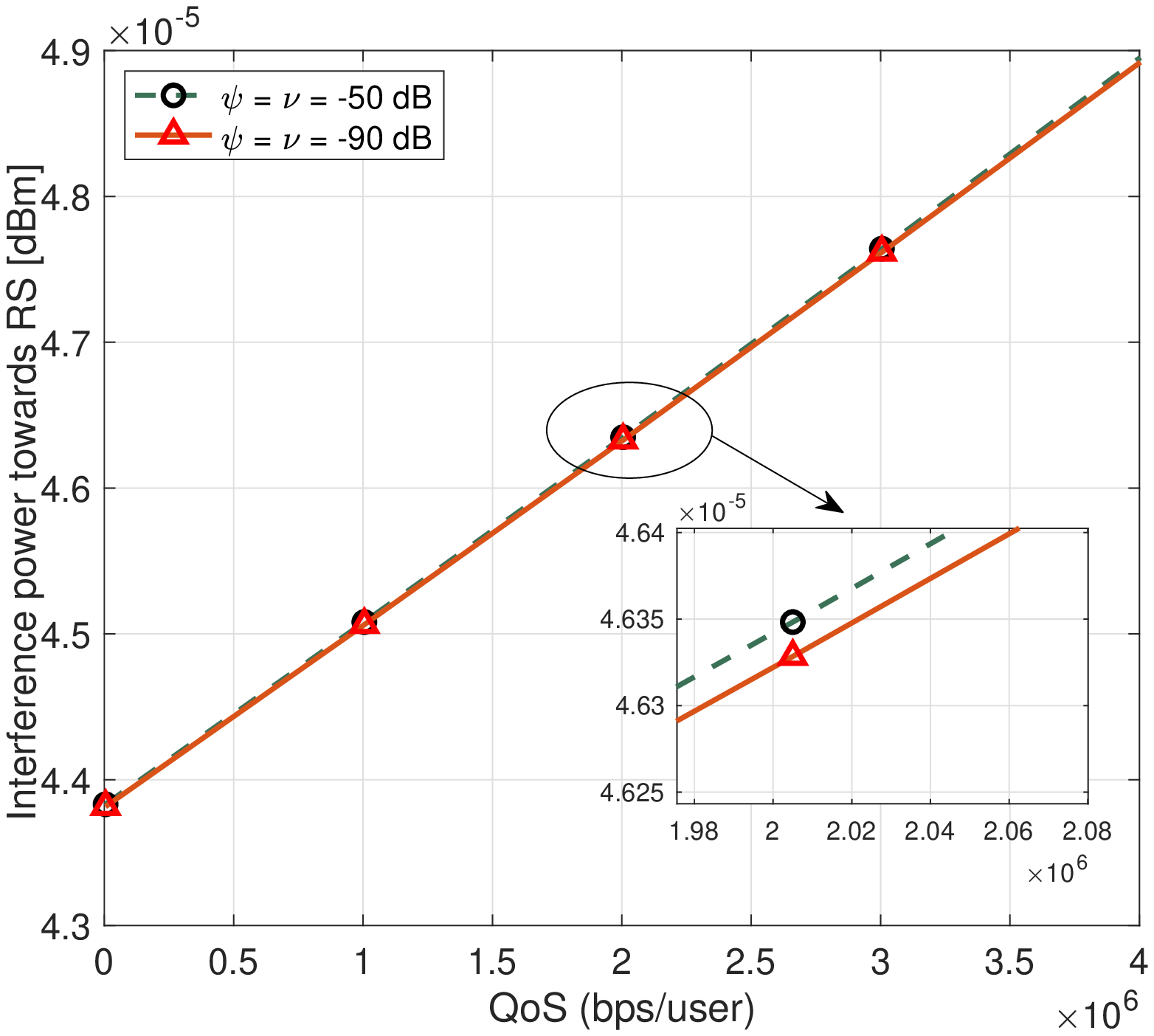}
  \captionof{figure}{Interference power towards RS vs. QoS per user at CS.}
  \label{sim_fig_6}
\end{minipage}
\end{figure*}
 
The estimated channel gain between the BS and $k$th UL user can be described as {$\tilde{\mathbf{H}}_{k}^{UL} = \sqrt{\wp_{k}^{UL}} \hat{\mathbf{H}}_{k}^{UL}$}, where $\hat{\mathbf{H}}_{k}^{UL}$ denotes small scale fading following a complex Gaussian distribution with zero mean and unit variance, and $\wp_{k}^{UL} = 10^{(-{\rm A}/10)},~A \in \{ {\rm LOS, NLOS} \}\footnote{\rm LOS=Line-of-sight and NLOS=Non-line-of-sight.}$ denotes the large scale fading consisting of path loss and shadowing. ${\rm LOS}$ and ${\rm NLOS}$ are computed based on a street canyon scenario \cite{Sun_2016}. The parameter $\alpha_c$ for LOS and NLOS are set as $2.0$ and $3.1$, respectively, while the value of shadow fading standard deviation $\sigma$ for LOS and NLOS are $2.9$ dB and $8.1$ dB, respectively. Similarly, we define the channels between UL users and DL users, between BS and DL users, between BS and RS, and between UL users and RS. To model the SI channel, the Rician model in~\cite{duarte2010} is adopted, wherein the SI channel is distributed as $\tilde{\mathbf{H}}_0 \sim \mathcal{CN}\left( \sqrt{\frac{K_R}{1+K_R}}\hat{\mathbf{H}}_0, \frac{1 }{1+K_R}\mathbf{I}_{N_0} \otimes \mathbf{I}_{M_0}\right)$, where  $K_R$ is the Rician factor and $\hat{\mathbf{H}}_0$ is a deterministic matrix\footnote{For simplicity, we take $K_R = 1$ and the matrix $\tilde{\mathbf{H}}_0$ of all ones for all simulations~\cite{nguyen2014}.}. Unless otherwise stated, we consider the following parameters for the CS and RS. For CS: thermal noise density $=-174$ dBm/Hz, noise figure at BS (users) $13(9)$ dB, $\tilde N=2$, $~\psi=\upsilon=-70$ dB, $\delta=\theta=0.1$, $R_{i,min}^{UL}=R_{j,min}^{DL}=5.0\times 10^5$ bps, $P_{i}=5$ dB, $P_{0}=10$ dB and CCI cancellation factor\footnote{It is essential to isolate UL and DL users in a FD system through smart channel assignments at a stage prior to the precoder/decoder design, so that the CCI is mitigated. This can be done by clustering the users into different groups through techniques, such as game theory, where the users with very strong CCI are not placed in the same group. In this work, the value of CCI cancellation factor represents the following: $0 \rightarrow100\%$ cancellation and $1\rightarrow 0\%$ cancellation.}$= 0.5$. For RS: $R=8$, $P_{FA}=10^{-5}$, velocity of target $=782$ knots, and distance of target from RS $=300$ m.
   \begin{figure*}[t!]
\centering
\begin{minipage}{.5\textwidth}
  \centering
  \includegraphics[width=1.027\linewidth]{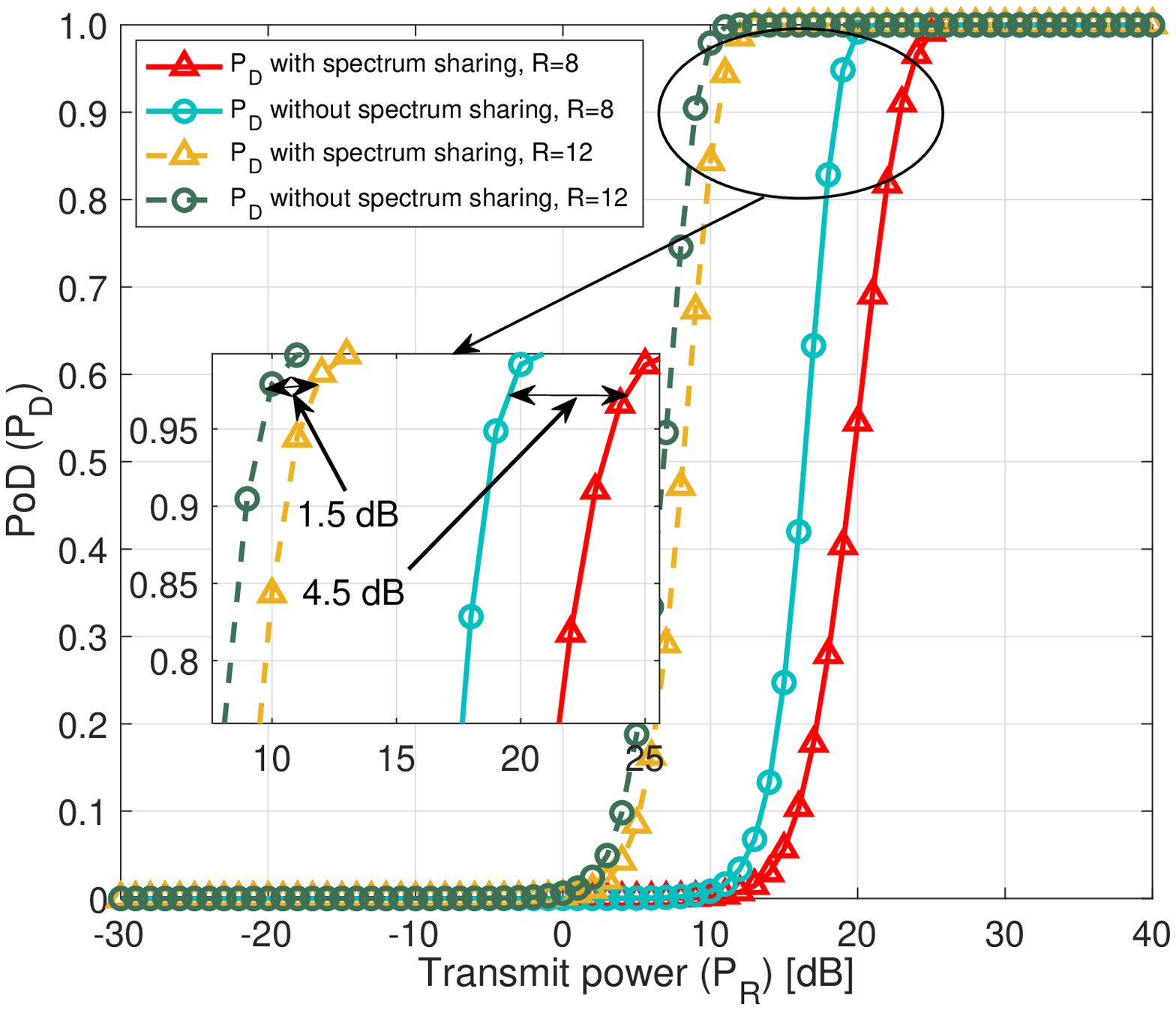}
  \captionof{figure}{{PoD of MIMO RS vs. MIMO RS's transmit power.}}
  \label{sim_fig_3}
\end{minipage}%
\begin{minipage}{.5\textwidth}
  \centering
  \includegraphics[width=1\linewidth]{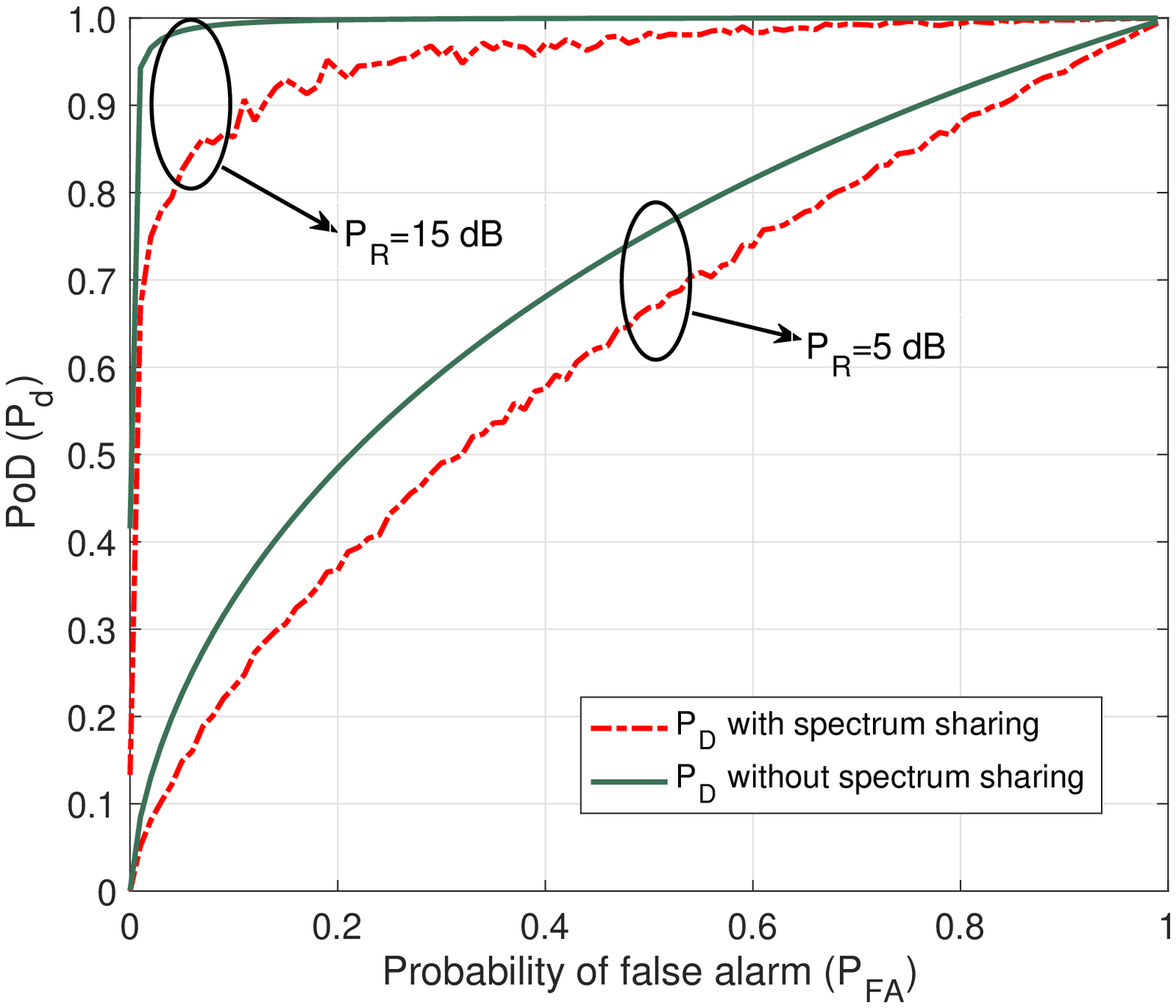}
  \captionof{figure}{PoD of MIMO RS vs. probability of false alarm at MIMO RS.}
  \label{sim_fig_4}
\end{minipage}
\end{figure*}
\subsubsection{Simulation Results}
To enable spectrum sharing, RS deploys Algorithm~1 to null the interference from $\text{RS}\rightarrow \text{CS}$, i.e., $\mathbf{W}_i\mathbf{ x}_R=\bold{0}\,,i\in\mathcal{S}$, which paves the way for the CS to use the RS's spectrum. This is simultaneously followed by the deployment of Algorithm~2 at CS, which maximises the PoD of RS by suppressing the interference from $\text{CS}\rightarrow \text{RS}$, while also providing QoS to its users. In the following examples, we illustrate the performance of both RS and CS under a spectrum sharing scenario utilising the proposed algorithms.  Due to the iterative nature of Algorithm~2, we begin by showing 1) its evolution in Fig.~\ref{sim_fig_1}, i.e., its convergence and 2) its complexity analysis in Fig.~\ref{sim_fig_2} in terms of complex multiplications required  with respect to (w.r.t) increasing number of antennas at CS and RS and users in the CS. It can be seen from Fig.~\ref{sim_fig_1} that the cost function, i.e., $I^{RAD}$ monotonically decreases and converges after $25-30$ iterations. Further, in Fig.~\ref{sim_fig_2}, the axes in red (left and bottom) represent the complexity w.r.t. number of users in the CS, while the axes in black (right and top) represent the complexity w.r.t. number of antennas at RS and the FD BS. It can be seen that the computational complexity of Algorithm 2 increases as the number of users or antennas are increased. Hence, it is imperative that the processing of Algorithm~2 be handled centrally at the FD BS, which inadvertently has high end computing capabilities. 

Next, we analyse the impact of spectrum sharing on the proposed two-tier model. In particular, we quantify the level of interference towards the RS for various levels of QoS that the CS can support by operating in FD mode. Accordingly, in Fig.~\ref{sim_fig_5} we show the interference power generated from the CS towards the RS as a function of transmitter/receiver $(\psi/\nu)$ distortion values for two different QoS requirements of the cellular users. Note that the transmitter/receiver distortion values reflect the amount of RSI left in the FD system. It can be seen from the figure that, as the RSI cancellation capability of the FD system increases, the interference power generated by the CS towards the RS decreases. This can be explained due to the fact that, when RSI is more, the CS has to use higher transmission power to overwhelm the RSI and maintain the QoS of the users, which results in more interference towards the RS. {More importantly, it can also be seen that the minimum guaranteed QoS for each user can be increased by a factor of $10\times$ or more if the RS increases its tolerance threshold for interference temperature by only $0.6$ dB.} Similarly, in Fig.~\ref{sim_fig_6} we show the effect of users' QoS for two different RSI values. It can be seen that as the QoS requirements of cellular users increases, the interference from CS towards RS  increases linearly. This can be explained due to the fact that, to provide higher QoS to the users, Algorithm~2 ensures transmission at higher power at the CS. Nevertheless, Algorithm~2 also ensures that for any specific QoS on the $x-$axis of Fig.~\ref{sim_fig_6}, the corresponding $y-$axis value represents the minimum interference that can be generated from CS towards the RS.

After quantifying the interference towards RS, we now determine its PoD. To detect a target in the far-field, the RS transmits NSP waveforms generated in Algorithm~1 and estimates parameters $\phi$ and $\alpha_r$ from the received signal that also includes $I^{RAD}$ (obtained from Algorithm~2). As a benchmark, we also simulate the scenario without spectrum sharing by generating orthogonal waveforms at the RS and setting $I^{RAD}=0$. This scenario relates to the case when the CS's BS is  unable to provide its users with any connectivity due to lack of spectrum resources. 

{Accordingly, in Fig.~\ref{sim_fig_3}, the PoD of the MIMO RS w.r.t. RS's transmit power is shown. Here, we consider two scenarios: 1) $R=8$ (straight lines) and 2) $R=12$ (dashed lines). It can be seen that for fixed $P_{FA}$, in order to achieve a particular $P_D$ the RS needs more power (to create NSP waveforms and withstand interference from CS to enable spectrum coexistence) than the case without spectrum sharing scenario. However, it can be seen that the RS needs more power when $R=8$ than $R=12$ to achieve similar performance. This is because, while the number of antennas at the CS (BS and users) are fixed, increasing the RS's antennas means that it has more degrees of freedom for reliable target detection and simultaneously nulling out interference towards the CS. This proves that large antenna arrays  can be used at the RS to facilitate spectrum sharing without any significant degradation in RS's performance.}

Finally, in Fig.~\ref{sim_fig_4}, we plot PoD for various $P_{FA}$ and $P_R$. Similar to the previous figure, the PoD of the RS is better at high $P_R$ and small $P_{FA}$ when the RS is not sharing its spectrum. However when both $P_R$ and $P_{FA}$ are small, PoD of RS for NSP waveforms is quite comparable to the case without spectrum sharing.

\section{Conclusion}\label{section_conclusion}
{{A two tier spectrum sharing framework was proposed, where 1) transceivers were jointly designed at a hardware impaired FD CS under imperfect CSI considerations 2) null-space based waveforms were designed at MIMO RS under perfect CSI considerations. In particular, the robust optimisation in the CS led to an intractable problem, which was transformed into an equivalent tractable semidefinite programming problem. Next, algorithms were proposed to suppress interference at both systems, thus maximising the PoD of RS and also maintaining a specific QoS for each user in CS. Finally, numerical results were used to demonstrated the effectiveness of the proposed algorithms, albeit certain trade-offs in RS's transmit power, PoD, and QoS of the users. In particular, it was seen that to facilitate spectrum sharing, thereby providing the users of CS with QoS of $5\times 10^{-5}$ bps/user, the MIMO RS needs to spend an extra power of $1.5-4.5$ dB depending on the number of antennas it uses. Overall, the designed framework provides the essential understanding for successful development of future cellular systems in conjunction with federal incumbents that can operate under same spectrum resources.}}
\appendices
\setcounter{equation}{0}
\renewcommand{\theequation}{A.\arabic{equation}}
{\section{Proof of Proposition \ref{NSP_prop}}\label{Appendix_A0}
In order to prove Proposition \ref{NSP_prop},  we first need to show that $\mathbf{P}$ is a projector. From \eqref{projection_matrix_eq1}, we have
\begin{align}\label{projection_matrix_eq2}
\mathbf{P}^H&=(\mathbf{X}\bold{\tilde\Omega}\mathbf{X}^H)^H=\mathbf{X}\bold{\tilde\Omega}^H\mathbf{X}^H=\mathbf{P}\,,\nonumber\\
\text{and} \,\,\mathbf{P}^2&=\mathbf{X}\bold{\tilde\Omega}\mathbf{X}^H\mathbf{X}\bold{\tilde{\Omega}}\mathbf{X}^H=\mathbf{P}.
\end{align}
The above equation holds due to the fact that $\mathbf{X}\mathbf{X}^H=\mathbf{I}$ as they are orthogonal matrices and $\bold{\tilde\Omega}^2=\bold{\tilde\Omega}$ by construction. 
Now in order to show that $\mathbf{P}$ is a projector, we show $\mathbf{P}\mathbf{x}=\mathbf{x}$, if $\mathbf{x}\in\text{range}(\mathbf{P})$. In other words, for some $\mathbf{w}$, $\mathbf{x}=\mathbf{P}\mathbf{w}$. Hence, from the above and \eqref{projection_matrix_eq2}, we have
\begin{align}
\mathbf{P}\mathbf{x}=\mathbf{P}(\mathbf{P}\mathbf{w})=\mathbf{P}^2\mathbf{w}=\mathbf{P}\mathbf{w}=\mathbf{x},\\
\mathbf{P}(\mathbf{P}\mathbf{x}-\mathbf{x})=\mathbf{P}^2\mathbf{x}-\mathbf{P}\mathbf{x}=\bold{0}.
\end{align}
Hence, $\mathbf{P}\mathbf{x}-\mathbf{x}\in\text{null}(\mathbf{P})$, which shows that $\mathbf{P}$ is a null-space projection matrix. Accordingly, 
\begin{equation}
\mathbf{W}\mathbf{P}^H=\mathbf{R}\bold{\bar\Omega}\mathbf{X}^H\mathbf{X}\bold{\tilde\Omega}\mathbf{X}^H=\bold{0}.
\end{equation}
which proves that $\mathbf P$ is an orthogonal projection matrix onto the null-space of $\mathbf W$.}
\setcounter{equation}{0}
\renewcommand{\theequation}{B.\arabic{equation}}
\section{Proof of Lemma \ref{Lemma_LB}}\label{Appendix_A1}
Since $\mathbf{A}\left(\phi\right){\textbf{P}}{\textbf{P}}^{H}\mathbf{A}^{H}\left(\phi\right)$ and $\hat{\boldsymbol\chi}$ are positive-definite, we have
\begin{align}\label{lb1}
\text{tr}\Big(\mathbf{A}\left(\phi\right){\textbf{P}}&{\textbf{P}}^{H}\mathbf{A}^{H}\left(\phi\right)\hat{\boldsymbol\chi}^{-1}\hat{\boldsymbol\chi}\Big)\,,\nonumber\\
&\leq\text{tr}\left(\mathbf{A}\left(\phi\right){\textbf{P}}{\textbf{P}}^{H}\mathbf{A}^{H}\left(\phi\right)\hat{\boldsymbol\chi}^{-1}\right)\text{tr}\left(\hat{\boldsymbol\chi}\right)\,.
\end{align}
Now, a lower bound for $\text{tr}\left(\mathbf{A}\left(\phi\right){\textbf{P}}{\textbf{P}}^{H}\mathbf{A}^{H}\left(\phi\right)\hat{\boldsymbol\chi}^{-1}\right)$ follows as
\begin{align}\label{lower_bound_fun}
\text{tr}\left(\mathbf{A}\left(\phi\right){\textbf{P}}{\textbf{P}}^{H}\mathbf{A}^{H}\left(\phi\right)\hat{\boldsymbol\chi}^{-1}\right)\geq\dfrac{\text{tr}\left(\mathbf{A}\left(\phi\right){\textbf{P}}{\textbf{P}}^{H}\mathbf{A}^{H}\left(\phi\right)\right)}{\text{tr}\left(\hat{\boldsymbol\chi}\right)}\,.
	\end{align}	
Using the property $\text{tr}(\textbf{A}_1+\textbf{B}_1)=\text{tr}(\textbf{A}_1)+\text{tr}(\textbf{B}_1)$, where $\textbf{A}_1$ and $\textbf{B}_1$ are square matrices and $\varphi=\text{tr}({\textbf{P}}{\textbf{P}}^{H}$), we can rewrite~\eqref{lower_bound_fun} under ${R}_{T}={R}_{R}={R}$ to obtain the desired result.

\setcounter{equation}{0}
\renewcommand{\theequation}{C.\arabic{equation}}
\section{Proof of Lemma \ref{Lemma_mse_vec_form}} \label{Appendix_A}
To prove this lemma, we first construct $\text{tr}\left(\mathbf{B}_{i}^{H}\mathbf{E}_{i}\mathbf{B}_{i}\right)$ using~\eqref{mse_matrix_mixed} as
\begin{align}\label{MSE_first_term}
\text{tr}\Big(\mathbf{B}_{i}^{H}&\mathbf{E}_{i}\mathbf{B}_{i}\Big)
	=\parallel \mathbf{B}_{i}^{H}\left(\mathbf{U}_{i}^{H} \mathbf{H}_{ii} \mathbf{V}_{i} -\mathbf{I}_{d_i}\right)\parallel_{F}^{2}\nonumber\\
	&+ \sum\nolimits_{j \in \mathcal{S},j\neq i}  \parallel \mathbf{B}_{i}^{H}\mathbf{U}_{i}^{H} \mathbf{H}_{ij} \mathbf{V}_{j}\parallel_{F}^{2}   \nonumber \\
	&+ {  \sum\nolimits_{j \in \mathcal{S}} \sum\nolimits_{\ell\in \mathcal{D}^{({T})}_j} \psi \parallel \mathbf{B}_{i}^{H}  \mathbf{U}_{i}^{H}  \mathbf{H}_{ij} \mathbf{\Xi}_{\ell} \mathbf{V}_{j}\parallel_{F}^{2}}\nonumber\\
	& +  \sum\nolimits_{j \in \mathcal{S}} \sum\nolimits_{\ell\in  \mathcal{D}^{({R})}_i}  \upsilon \parallel \mathbf{B}_{i}^{H}\mathbf{U}_{i}^{H} \mathbf{\Xi}_{\ell} \mathbf{H}_{ij} \mathbf{V}_{j}\parallel_{F}^{2}\nonumber\\
	&+P_{R}\parallel \mathbf{B}_{i}^{H}\mathbf{U}_{i}^{H} \textbf{W}_{i}\parallel + \sigma_i^2 \parallel \mathbf{B}_{i}^{H} \mathbf{U}_{i}^{H} \parallel_{F}^{2},	
\end{align}  
where $\mathcal{D}^{({R})}_j$ and $\mathcal{D}^{({T})}_j$ denote the set $\{1 \cdots \tilde{N}_j \}$ and $\{1 \cdots \tilde{M}_j \}$, respectively, while $\mathbf{\Xi}_{\ell}$ represents a square matrix with zero elements, except for the $\ell$-th diagonal element, which is equal to $1$. Utilising the $\mathrm{vec}(\cdot)$ operation, and the identity $\|\mathrm{vec}\left(\mathbf{A} \right)\|_2^2= \text{tr} \left\{\mathbf{A} \mathbf{A}^H \right\}$,~\eqref{MSE_first_term} from above can be reformulated as
\begin{align} \label{MSE_vec}
\text{tr}\Big(\mathbf{B}_{i}^{H}&\mathbf{E}_{i}\mathbf{B}_{i}\Big)=\left\| \mathrm{vec} \left(\mathbf{B}_{i}^{H}\left(\mathbf{U}_{i}^{H} \mathbf{H}_{ii} \mathbf{V}_{i}-\mathbf{I}_{d_i} \right)\right)\right\|_{2}^{2} \nonumber\\
&+  \sum_{j \in \mathcal{S},j\neq i} \left\|\mathrm{vec}\left(\mathbf{B}_{i}^{H} \mathbf{U}_i^H \mathbf{H}_{ij}\mathbf{V}_j\right) \right\|_2^2 	 \nonumber \\
&+\sum\limits_{j \in \mathcal{S}} \sum\limits_{\ell\in \mathcal{D}^{({T})}_j} \psi \left\|\mathrm{vec}\left(\mathbf{B}_{i}^{H} \mathbf{U}_i^H \mathbf{H}_{ij} \mathbf{\Xi}_{\ell} \mathbf{V}_j  \right)\right\|_2^2  \nonumber\\
&+  \sum\limits_{j \in \mathcal{S}} \sum\limits_{\ell\in \mathcal{D}^{({R})}_i} \upsilon \left\|\mathrm{vec}\left(\mathbf{B}_{i}^{H} \mathbf{U}_i^H \mathbf{\Xi}_{\ell}  \mathbf{H}_{ij} \mathbf{V}_j  \right)\right\|_2^2\\
&+P_{R}\left\|\mathrm{vec} \left(\mathbf{B}_{i}^{H}\mathbf{U}_{i}^{H} \textbf{W}_{i}\right)\right\|_2^2 +  \sigma_i^2\left\|\mathrm{vec} \left(\mathbf{B}_{i}^{H}\mathbf{U}_{i}^{H}\right) \right\|_2^2\,.\nonumber
\end{align}  
By utilising the identity $\mathrm{vec}(\mathbf{ABC})= \left(\mathbf{C}^T\otimes \mathbf{A} \right)\mathrm{vec}\left(\mathbf{B}\right)$,~\eqref{MSE_vec} in the above can be rewritten as $ \left\| \textbf{z}_{ij}  \right\|_2^2$, where $\textbf{z}_{ij}$ is defined as
{ 	 \begin{align} \label{variables1}
\textbf{z}_{ij}\!=\!\! \left[\! \begin{array}{c} 
					\mathrm{vec} \left(\mathbf{B}_{i}^{H}\left(\mathbf{U}_{i}^{H} {\mathbf{H}}_{ii} \mathbf{V}_{i}-\mathbf{I}_{d_i} \right)\right) \\
					\left\lfloor\left( \mathbf{V}_j^T \otimes \left(\mathbf{B}_{i}^{H}\mathbf{U}_i^{H}\right) \right)\mathrm{vec} \left({\mathbf{H}}_{ij}\right)\right\rfloor_{j \in \mathcal{S},j\neq i}   \\
					\!\!\! \left\lfloor \left\lfloor \sqrt{\psi}\left( (\mathbf{\Xi}_{\ell} \mathbf{V}_{j})^T \otimes\left(\mathbf{B}_{i}^{H} \mathbf{U}_{i}^{H}\right) \right) \mathrm{vec} \left({\mathbf{H}}_{ij}\right) \right\rfloor_{\ell \in \mathcal{D}^{({T})}_j} \right \rfloor_{j \in \mathcal{S}}\!\!   \\
					\!\!\! \left\lfloor \left\lfloor  \sqrt{\upsilon} \left( \mathbf{V}_j^T \otimes \left(\mathbf{B}_{i}^{H}\mathbf{U}_{i}^{H} \mathbf{\Xi}_{\ell}\right)\right) \mathrm{vec} \left({\mathbf{H}}_{ij}\right) \right\rfloor_{\ell \in \mathcal{D}^{({R})}_i}   \right\rfloor_{j \in \mathcal{S}} \!\!  \\
					\!\!\! P_{R}\left(\mathbf{I}_{R}^{T}\otimes \left(\mathbf{B}_{i}^{H}\mathbf{U}_{i}^{H}\right)\right)\mathrm{vec} \left(\textbf{W}_{i}\right)\!\!  \\
					\sigma_i \mathrm{vec} \left(\mathbf{B}_{i}^{H}\mathbf{U}_{i}^{H}\right)
				\end{array}\!\!
				\right]\,
\end{align}     }
In a similar way, we can also represent $I^{RAD}$ into vector form as
					 \begin{align}\label{int_pack}
						I^{RAD}= &\sum\nolimits_{j \in \mathcal{S}} \left(\left\|\mathrm{vec}\left(\mathbf{G}_{lj} \mathbf{V}_j\right)\right\|_2^2  \right.\nonumber\\
						&\left.+ \sum\nolimits_{\ell \in \mathcal{D}^{({T})}_j}   \psi \left\| \mathrm{vec}\left(\mathbf{G}_{lj} \mathbf{\Xi}_{\ell} \mathbf{V}_{j} \right)  \right\|_2^2\right)\,.
					\end{align}    
					Now, applying the same identity $\mathrm{vec}(\mathbf{ABC})= \left(\mathbf{C}^T\otimes \mathbf{A} \right)\mathrm{vec}\left(\mathbf{B}\right)$,~\eqref{int_pack} in above can be constructed as $\left\|  \boldsymbol{\iota} \right\|_2^2$, where $ \boldsymbol{\iota}$ is defined as 	
				 \begin{align}	\boldsymbol{\iota} \!=\!\left[ \!\!
			\begin{array}{c}
				\!	\left\lfloor  \left( \mathbf{V}_j^T \otimes \mathbf{I}_{T} \right)\mathrm{vec} \left(\mathbf{G}_{j}\right) \right\rfloor_{j \in \mathcal{S}}  \\
				\!\sqrt{\psi}\left\lfloor \left\lfloor  \left( (\mathbf{\Xi}_{\ell} \mathbf{V}_{j})^T \otimes \mathbf{I}_{T} \right) \mathrm{vec} \left(\mathbf{G}_{j}\right)  \right \rfloor_{\ell \in \mathcal{D}^{(T)}_j}  \right\rfloor_{j \in \mathcal{S}}  
			\end{array} \!\!\right].
			\end{align}  
	
\setcounter{equation}{0}
\renewcommand{\theequation}{D.\arabic{equation}}
\section{Useful Lemmas}\label{Appendix_B}
		\begin{lem}   \label{lemma5}
		If $\mathbf{P,~Q,~A}$ are matrices with $\mathbf{A}$ = $\mathbf{A}^H$, then semi-infinite LMI of the form of $\mathbf{A} \succeq \mathbf{P}^H\mathbf{XQ} + \mathbf{Q}^H\mathbf{X}^H\mathbf{P},\forall \mathbf{X} : \lVert \mathbf{X}\rVert_F \leq \rho, $
			holds \textit{iff}  $ \exists \epsilon \geq 0$ such that
			 \begin{eqnarray}
			\left[
			\begin{array}{cc}
			\mathbf{A}-\epsilon\mathbf{Q}^H\mathbf{Q} & -\rho\mathbf{P}^H \\
			-\rho\mathbf{P} & \epsilon\mathbf{I}
			\end{array} \right] \succeq 0.
			\end{eqnarray}    
		\end{lem}
	
	\begin{lem}   \label{lemma6}
	Schur Complement Lemma \cite{Boyd04}: Let $\mathbf{Q}$ and $\mathbf{R}$ are symmetric matrices. Then
			 \begin{eqnarray}
				\left[\begin{array}{cc}
					\mathbf{Q} & \mathbf{S} \\
					\mathbf{S}^* & \mathbf{R}
				\end{array}\right] \succeq \mathbf{0} \quad \triangleq \quad \mathbf{R} \succeq \mathbf{0},~ \mathbf{Q}-\mathbf{S}\mathbf{R}^{-1}\mathbf{S}^* \succeq \mathbf{0}.
			\end{eqnarray}  
			        \end{lem}
\balance		        

\end{document}